\documentclass[conference]{IEEEtran}
\newcommand{\myparagraph}[1]{\vspace{0.3\baselineskip}\noindent{\textbf{#1.}}~}

\usepackage[linesnumbered,ruled,vlined,boxed,noend]{algorithm2e}
\usepackage{booktabs}
\usepackage{amsmath}
\usepackage{amsthm}
\usepackage{amssymb}

\usepackage[normalem]{ulem} 
\usepackage{xcolor}
\usepackage{soul}
\setul{0.5ex}{0.3pt}
\setulcolor{green} 

\usepackage{graphicx}
\usepackage{enumitem}
\usepackage{multirow}
\usepackage{colortbl}
\usepackage{diagbox}
\usepackage{tabularray}
\usepackage{dsfont} 
\usepackage{mathrsfs}
\usepackage{algorithmic}
\usepackage{amssymb}

\setlist[itemize]{leftmargin=*}

\usepackage[colorlinks=true,linkcolor=red,urlcolor=black,citecolor=blue]{hyperref}

\newcommand{\bfs}{{{\texttt{BFS}}}\xspace}
\newcommand{\dfs}{{{\texttt{DFS}}}\xspace}
\newcommand{\ball}{{{\texttt{MBB}}}\xspace}
\newcommand{\rectangle}{{{\texttt{MBR}}}\xspace}
\newcommand{\blb}{\mathbf{bp}}
\newcommand{\bub}{\mathbf{tp}}
\newcommand{\bq}{\mathbf{q}}
\newcommand{\bp}{\mathbf{p}}
\newcommand{\bo}{\mathbf{o}}

\newcommand{\BDLtree}{{{\texttt{BDL-tree}}}\xspace}
\newcommand{\Ball}{{{\texttt{Ball-tree}}}\xspace}

\newcommand\blfootnote[1]{%
  \begingroup
  \renewcommand\thefootnote{}\footnote{#1}%
  \addtocounter{footnote}{-1}%
  \endgroup
}

\newcommand{\kdtree}{{{\texttt{BMKD-tree}}}\xspace}

\newcommand{\aitree}{{{``\texttt{AI+R}''\texttt{-tree}}}\xspace}

\newcommand{\Unik}{{{\texttt{Unik}}}\xspace}
\newcommand{\TRIO}{{{\texttt{TRIO}}}\xspace}

\newcommand{\Elkan}{{{\texttt{Elkan}}}\xspace}

\newcommand{\finding}{\noindent{\underline{\textit{Observations}}}.~}

\newcommand{\learn}{\texttt{UnIS}\xspace}

\newcommand{\ra}[1]{\renewcommand{\arraystretch}{#1}}

\newcommand{\Tdrive}{{{\texttt{T-drive}}}\xspace}
\newcommand{\Porto}{{{\texttt{Porto}}}\xspace}
\newcommand{\Shapenet}{{{\texttt{Shapenet}}}\xspace}

\newcommand{\POI}{{{\texttt{ArgoPOI}}}\xspace}
\newcommand{\AVL}{{{\texttt{ArgoAVL}}}\xspace}
\newcommand{\PC}{{{\texttt{ArgoPC}}}\xspace}
\newcommand{\Trajectory}{{{\texttt{ArgoTraj}}}\xspace}
\newcommand{\Apollo}{{\texttt{Apollo}}\xspace}
\newcommand{\UnIS}{{{\texttt{UnIS}}}\xspace}

\usepackage{etoolbox}
\usepackage{marginnote}
\makeatletter
\let\oldmarginnote\marginnote
\renewcommand*{\marginnote}[1]{%
  \begingroup%
  \ifodd\value{page}
    \if@firstcolumn\normalmarginpar\fi
  \else
    \if@firstcolumn\else\normalmarginpar\fi
  \fi
  \oldmarginnote{\hspace{5pt}\textcolor{brown}{#1}}%
  \endgroup%
}
\usepackage{xcolor}

\newtheorem{definition}{Definition}

\newtheorem{lemma}{Lemma}
\newcommand{\algrule}[1][.2pt]{\par\vskip.5\baselineskip\hrule height #1\par\vskip.5\baselineskip}

\newcounter{cN}
\definecolor{Maroon}{RGB}{128, 0, 0}

\definecolor{NavyBlue}{rgb}{0.0, 0.0, 0.5}

\ifCLASSOPTIONcompsoc
  \usepackage[nocompress]{cite}
\else
  \usepackage{cite}
\fi
\ifCLASSINFOpdf
\else
\fi  
\begin{document}
\title{Updatable Balanced Index for Fast On-device Search with Auto-selection Model}


\author{\IEEEauthorblockN{Yushuai Ji\IEEEauthorrefmark{4}, Sheng Wang\IEEEauthorrefmark{4}\IEEEauthorrefmark{1}, Zhiyu Chen\IEEEauthorrefmark{2}, Yuan Sun\IEEEauthorrefmark{3}, and Zhiyong Peng\IEEEauthorrefmark{4}}
\IEEEauthorblockA{\IEEEauthorrefmark{4}School of Computer Science, Wuhan University\\  \IEEEauthorrefmark{2}Amazon.com, Inc.\\ \IEEEauthorrefmark{3}La Trobe Business School, La Trobe University}
\IEEEauthorblockA{[yushuai, swangcs, peng]@whu.edu.cn, zhiyuche@amazon.com, yuan.sun@latrobe.edu.au}
}









\markboth{IEEE TRANSACTIONS ON KNOWLEDGE AND DATA ENGINEERING,~Vol.~X, No.~X, MAY~2024}%
{Shell \MakeLowercase{\textit{et al.}}: Bare Demo of IEEEtran.cls for Computer Society Journals}
\maketitle
\begin{abstract} 
Diverse types of edge data, such as 2D geo-locations and 3D point clouds, are collected by sensors like lidar and GPS receivers on edge devices. 
On-device searches, such as $k$-nearest neighbor ($k$NN) search and radius search, are commonly used to enable fast analytics and learning technologies, such as $\mathrm{k}$-means dataset simplification using $k$NN.
To maintain high search efficiency, a representative approach is to utilize a balanced multi-way KD-tree (BMKD-tree).
However, the index has shown limited gains, mainly due to substantial construction overhead, inflexibility to real-time insertion, and inconsistent query performance.
In this paper, we propose \learn to address the above limitations.
We first accelerate the construction process of the BMKD-tree by utilizing the dataset distribution to predict the splitting hyperplanes.
To make the continuously generated data searchable, we propose a selective sub-tree rebuilding scheme to accelerate rebalancing during insertion by reducing the number of data points involved.
We then propose an auto-selection model to improve query performance by automatically selecting the optimal search strategy among multiple strategies for an arbitrary query task. Experimental results show that \learn achieves average speedups of 17.96× in index construction, 1.60× in insertion, 7.15× in $k$NN search, and 1.09× in radius search compared to the BMKD-tree.
We further verify its effectiveness in accelerating dataset simplification on edge devices, achieving a speedup of 217× over Lloyd's algorithm.


\end{abstract}

\IEEEdisplaynontitleabstractindextext
\IEEEpeerreviewmaketitle

\section{Introduction}
\label{sec:intro}
Edge devices such as autonomous vehicles \cite{WangBCC21} and drones \cite{Xiao24} are typically equipped with sensors like GPS and lidar. These sensors continuously generate large volumes of data points, such as 2D geo-location from GPS and 3D point clouds \cite{Chang2019} from lidar. The on-device search, such as $k$NN search and radius search, can be applied to such data for efficient analysis and learning \cite{IP22, MatthiesHCJ08, Xu2023, Zhu22}. For example, $k$NN search is the core idea behind several clustering accelerators \cite{Yushuai2024, Yushuai2025}, which can improve $\mathrm{k}$-means efficiency. 
\blfootnote{\IEEEauthorrefmark{1}Sheng Wang is the corresponding author.}

The most popular method to support on-device search is through the use of indices. Specifically, an index is a data structure that improves search efficiency at the expense of additional storage space. In general, the indexes that are widely used can be classified into three categories \cite{James2023}: 1) table-based indexes, 2) graph-based indexes, and 3) tree-based indexes. However, table-based and graph-based indexes cannot improve on-device search efficiency, as both inherently trade off result completeness for efficiency, and ensuring exactness requires traversing all data points. Tree-based indexes can avoid unnecessary computations when searching for exact results by assigning similar data points to the same sub-space. For example, given a query point $\mathbf{p}^*$, $k$NN will traverse the tree iteratively from the root node and apply the \textit{triangle inequality} \cite{friedman1977algorithm} to prune the sub-spaces that are far from $\mathbf{p}^*$. Once the traversal on the unpruned sub-spaces is done, the exact results will be returned.

However, when dealing with large-scale datasets, tree-based indexes often struggle to efficiently retrieve up-to-date data for supporting real-time downstream analytics and learning \cite{Yushuai2024,ZhuF22}. For example, indexes such as balanced multi-way KD-trees (BMKD-trees) \cite{ProcopiucAAV03} can accelerate the updatable $k$-means process \cite{Yushuai2024} by improving the efficiency of $k$NN search.
Yet in practice, when processing large-scale datasets—for instance, clustering 5 million data points from the \PC \cite{argoverse}—these accelerators, such as Dask-means \cite{Yushuai2024}, achieve only a 1.07$\times$ average speedup over the traditional $k$-means \cite{Lloyd1982}. The reduced acceleration is due to the substantial indexing overhead, limited support for real-time insertion, and inconsistent query performance. Details are provided below.


\begin{figure}
	\centering
\includegraphics[width=0.47\textwidth]{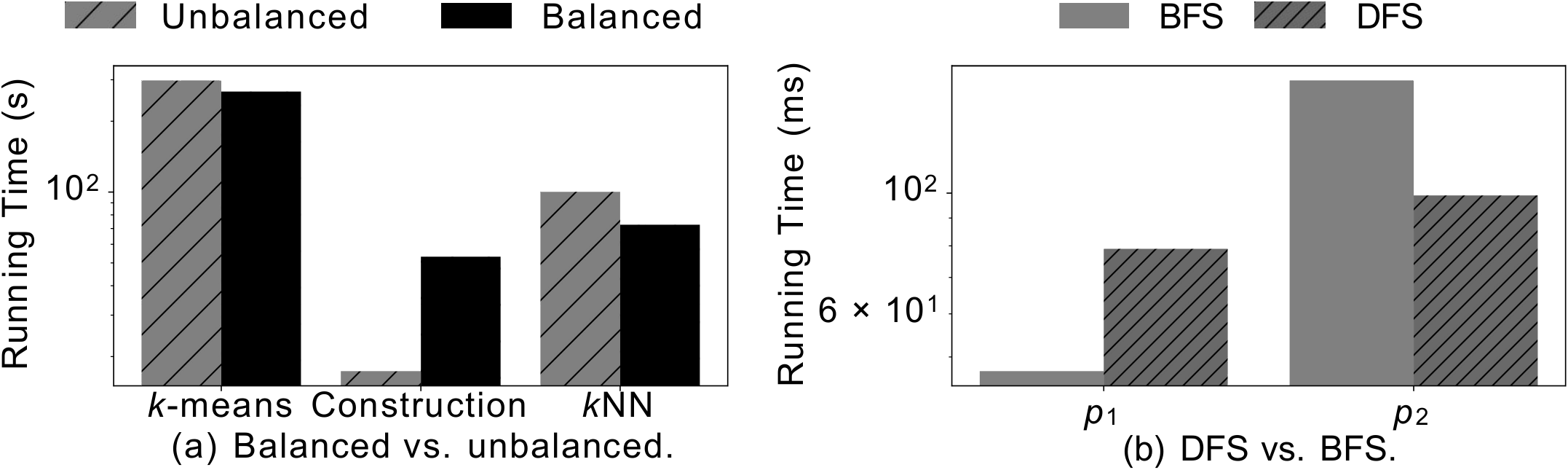}
	\vspace{-0.5em}
	\caption{Performance comparisons of the balanced multi-way KD-tree.}
	\label{fig:intro}
	\vspace{-1.5em}
\end{figure}

\myparagraph{Time-Intensive Construction Process} Constructing a BMKD-tree can reduce the average search path length, which is critical for improving query performance. However, this process requires sorting to compute splitting hyperplanes that divide the dataset into equally sized subspaces, which is time-consuming \cite{Brown2014}. As shown in Fig.~\ref{fig:intro}(a), the BMKD-tree outperforms the unbalanced one in $k$NN. Nonetheless, its construction takes more than four times longer. Hence, although $k$NN search using BMKD-trees can accelerate the $\mathrm{k}$-means algorithm more effectively, the time saved in searching is offset by the high index construction cost.

\myparagraph{Rebuilding Overhead} Updating-intensive operations disrupt the structure of BMKD-trees, hindering efficient access to the latest data points \cite{XuLLXCZLYYYCY23}.
The tree periodically rebuilds to ensure that queries efficiently retrieve the most recent data points, but the rebuilding process is time-consuming.
A classical method to improve rebuild efficiency by using the \textit{scapegoat strategy}~\cite{yixi2021}, which involves the parallel rebuilding of sub-trees to achieve balance. However, when facing a large volume of real-time insertions, sub-tree rebuilding occurs more frequently, making the strategy less effective.

\myparagraph{Inconsistent Query Performance} No single search strategy can perform well for all queries. For example, in Fig.~\ref{fig:intro}(b), we perform $k$NN queries with two points, $\mathbf{p}_1$ and $\mathbf{p}_2$. We use two traversal strategies to explore the search space: Breadth-First Search (BFS) and Depth-First Search (DFS). The DFS-based strategy is more efficient for $\mathbf{p}_1$, while the BFS-based strategy is better for $\mathbf{p}_2$. This is because the dimensionality and distribution of the dataset have varying effects on the DFS and BFS search efficiency \cite{SternKFH10}. Manual selection of optimal strategies is impractical due to the lack of a simple and discernible rule \cite{Wolpert1997}. A feasible solution to improve the efficiency of selecting optimal strategies is using auto-selection models \cite{SONG2012}. However, existing models \cite{Abdullah2022, Wang2020} encounter inefficiencies during sample generation, due to the absence of a fast and informative feature extraction method.

To address these challenges, we propose \UnIS, an \underline{u}pdatable bala\underline{n}ced \underline{i}ndex for fast on-device \underline{s}earch with an auto-selection model. It is designed to accelerate both search and update operations for edge data over the BMKD-tree. Overall, we make the following contributions in this paper:

\begin{itemize}
  \item To accelerate the construction of the BMKD-tree, we leverage data distribution to predict splitting hyperplanes instead of performing sorting, thereby improving the efficiency of space partitioning (see Section~\ref{sec:Construction}).

  \item We propose a selective sub-tree rebuilding scheme that identifies and reconstructs unbalanced sub-tree portions, thereby accelerating index rebalancing and improving updating efficiency (see Section~\ref{sec:Insertion}).

  \item To achieve fast exact searches for arbitrary query tasks, we design a lightweight auto-selection model to predict the optimal search strategy from a set of candidate strategies for each query task over the BMKD-tree (see Section~\ref{sec:auto}).
  
  \item Experiments show that \learn achieves average speedups of 17.96× in index construction, 1.60× in insertion, 7.51× in $k$NN search, and 1.09× in radius search. We also verify its effectiveness in accelerating $\mathrm{k}$-means on edge devices, achieving an average speedup of 217× (see Section~\ref{sec:exp}).
\end{itemize}
\section{Related Work}
\label{sec:related_sec}
\myparagraph{KD-tree and Its Variants}
The traditional method of constructing a KD-tree \cite{bentley1975} is to recursively generate a splitting hyperplane by averaging the maximum and minimum values in a specific dimension to split the space into two. However, using the average does not partition the dataset evenly, potentially leading to high vector volume in certain sub-spaces. Hence, the sub-spaces require more partitions compared to those with fewer vectors, resulting in increased tree depth (longer average search path length 
\cite{BereczkyDNR16}) and impairing query performance. 

Instead of using means, Friedman et al. \cite{friedman1977algorithm} propose using medians to construct a balanced KD-tree to partition the dataset into sub-spaces evenly, ensuring that leaf nodes reside at the same tree depth level. However, the traditional median-finding scheme involves sorting \cite{Brown2014,friedman1977algorithm}, which is known for its time-intensive nature. To address the limitation, Brown \cite{Brown2014} pre-sorts the data in each dimension and hence decreases the sorting time in construction. Blelloch et al. \cite{BlellochD22} reduce the sorting time by applying the well-known \textit{Z-order} technique, which reduces vectors' dimensionality, to improve construction efficiency.
However, as dimensionality increases, the sorting time grows significantly, leading to higher time costs, rendering both methods ineffective in accelerating index construction beyond four dimensions.

To further improve the query performance, Procopiuc et al. \cite{ProcopiucAAV03, BereczkyDNR16} utilize the idea of the median-finding scheme within the multi-way KD-tree, which splits space into more than two sub-spaces to reduce the search path length on average. The index determines an appropriate partition number through experimentation \cite{BereczkyDNR16}, testing different values of partition number to find the one that achieves the optimal query performance. However, sorting is still conducted to locate the equidistant points for space partitioning after determining the partition number. Yesantharao et al. \cite{BDLtree2021} partition the dataset into subsets and construct a balanced multi-way KD-tree for each subset, thereby reducing the sorting time as each subset is sorted individually, yet sorting is still required.

Recently, several studies~\cite{GuFCL0W23, Lan2023} have utilized learning techniques \cite{kraska2018case} on tree-structured indexes, which use machine learning (ML) models to capture the relationship of the position of vectors through the 
Cumulative Distribution Function (CDF) of the dataset. However, training ML models \cite{CaiCXWXZ23, Abdullah2022} to meet exact search requirements takes a substantial amount of time before they positively impact query performance.

\myparagraph{Index Insertion and Rebalance} To ensure efficient retrieval of continuously generated vectors, existing indexes periodically rebalance themselves during vector insertion \cite{XuLLXCZLYYYCY23}, utilizing either out-of-place or in-place insertion technologies. Out-of-place insertion \cite{WeiWWLZ0C20, Wang21} periodically merges delta indexes (rebuilt by the inserted vectors) into the base index via a global rebuilding process. However, the process requires additional memory to construct a duplicate index, which replaces the unbalanced one once the index is fully constructed.

In contrast, in-place insertion \cite{XuLLXCZLYYYCY23} involves inserting vectors into the existing index with dynamic rebalancing. Current indexes using in-place insertion include tree-structured indices \cite{yixi2021, XuLLXCZLYYYCY23} and cluster-based indices \cite{Jie2019, WeiWWLZ0C20}. Notably, cluster-based indices replace the node of the tree-structured index with clusters and thus can be regarded as a type of tree-structured index. Tree-structured index rebuilding typically employs the scapegoat strategy \cite{galperin1993scapegoat, yixi2021, BDLtree2021} to accelerate the rebalancing process by reconstructing sub-trees rather than fully rebuilding the entire tree. When facing update-intensive tasks, the tree-structured indexes undergo frequent partial rebuilds, resulting in the strategy being unable to reduce the number of vectors involved in rebuilding compared to global rebuilding.
\begin{table}[t]
\centering
\setlength{\tabcolsep}{10pt}
\caption{Summary of notations.}
\label{tab:notations}
\vspace{-1em}
\scalebox{1.04}{\begin{tabular}{cc}  
\toprule   
\textbf{Notation} & \textbf{Description}  \\  
\midrule
$n \in \mathbb{Z}^{+}$ &  The dataset size. \\
$k \in \mathbb{Z}^{+}$ &  The $k$ nearest neighbors. \\
$\mathrm{k} \in \mathbb{Z}^{+}$ &  The $\mathrm{k}$ clusters. \\
$\mathbf{p} =(x_1,x_2,...,x_d) \in \mathbb{R}^{d}$ &  The vector.\\
$\mathbf{D}=\{\mathbf{p}_i\}_{i=1}^{n} \in \mathbb{R}^{n \times d}$ &  The dataset.\\
$T$ &  The tree-structured index. \\
$c \in \mathbb{Z}^{+}$ &  The leaf node capacity. \\
$\delta \in (0,1)$ &  The sampling rate. \\
$d \in \mathbb{Z}^{+}$ &  The vector data dimension. \\
$\mathbf{P}\subseteq \mathbf{D}$ & The vector subset.\\
$\mathbf{O}$, $\mathbf{o} \in \mathbb{R}^{d}$ &  The pivot set and the pivot.\\
$\mathbf{M}$ &  The recursive CDF model. \\
\bottomrule  
\end{tabular}}
\vspace{-1.7em}
\end{table}

\myparagraph{Optimal Search Strategy Selection} The lack of a discernible rule makes it challenging to determine the optimal search strategy for an arbitrary query \cite{Wolpert1997}. An effective method to address this limitation is the auto-selection model \cite{Shapira2021}, which trains the classifier using meta-knowledge (or meta-features) extracted from previous similar tasks to identify the top-performing methods for unseen tasks.

Most auto-selection models \cite{SONG2012, Abdullah2022, Wang2020} aim to find the fastest or most accurate method for a given task. For example, \Unik \cite{Wang2020} models the fastest $\mathrm{k}$-means algorithm selection as a multi-label classification problem. It extracts features and trains a classifier from samples to predict the optimal algorithm pipeline for accelerating $\mathrm{k}$-means clustering tasks. However, \Unik only extracts basic features for each task, such as the data volume for the dataset, which cannot achieve a high prediction accuracy for arbitrary datasets. \TRIO \cite{Shapira2021} extracts more features to improve prediction accuracy, yet it overlooks the time costs, for instance, those incurred by utilizing the graph convolutional neural network \cite{li2018adaptive} for feature extraction. 

\myparagraph{Remarks} We can observe that: 1) the balanced KD-tree and its variants, such as BMKD-tree, can provide fast queries, but they still consume too much time on construction; 2) all insertion methods discussed necessitate significant time for index reconstruction to keep balance, thereby ensuring efficient queries; 3) existing auto-selection models suffer from inefficiencies in feature extraction due to the absence of a fast and informative feature extraction method.
\section{Preliminaries}

\subsection{Notations}
We use different text formatting styles to represent mathematical concepts: plain letters for scalars, bold letters for vectors, capitalized letters for objects, and bold capitalized letters for a set containing data points. 
For example, $x$ stands for a scalar, $\mathbf{p}$ represents a data point, $N$ denotes a node, and $\mathbf{D}$ represents a dataset. Without loss of generality, we denote the $d$-dimensional Euclidean space as $\mathbb{R}^{d}$, the set of positive real numbers as $\mathbb{R}^+$, and the set of positive integers as $\mathbb{Z}^{+}$. We represent the set of scalar values from the $j_{1}$-th to $j_{2}$-th elements in the $i$-th dimension of $\mathbf{D}$ as $\{x_{ij}\}_{i=j_1}^{j_2}$. The vector length is denoted by $|\cdot|$, and $\text{dist}(\cdot)$ stands for the Euclidean distance measure. The notation details are presented in Table~\ref{tab:notations}.
\begin{figure}
	\centering
 \includegraphics[width=.47\textwidth]{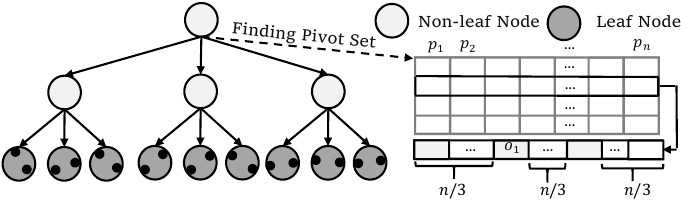}
	\vspace{-0.7em}
	\caption{A balanced multi-way KD-tree with a partition number of $t=3$.}
	\label{fig:multi-kd}
	\vspace{-1em}
\end{figure}

\subsection{Balanced Multi-way KD-tree}
The balanced multi-way KD-tree \cite{ProcopiucAAV03, BDLtree2021} consists of pivot sets, non-leaf nodes, and leaf nodes, as outlined below:

\begin{definition}(\textbf{Pivot Set}) 
  \label{def:pivot}
  Given a dataset $\mathbf{D} = \{\mathbf{p}_i\}_{i=1}^{n} \in \mathbb{R}^{n \times d}$, the partition number denoted as $t$, and a specified dimension symbolized by $j$, a pivot set consists of ranked values denoted by $\mathbf{O}=\{\mathbf{o}_i\}_{i=1}^{t-1} \in \mathbb{R}^{t-1}$, where $o_i$ corresponds to the value associated with the $\frac{i}{t}$-th percentile of $\{x_{ij}\}_{i=0}^{n}$. 
\end{definition}

\begin{definition}(\textbf{Non-leaf Node}) 
  \label{def:node1}
   A non-leaf node is defined as $N=(\mathbf{O},\mathbf{C}_{nl},j)$, where $\mathbf{O}$ represents the pivot set, $\mathbf{C}_{nl}$ is the children of $N$, and $j$ stands for the specific dimension. Here, $\mathbf{C}_{nl}[i]$ denotes the $i$-th sub-tree rooted at $N$.
\end{definition}

\begin{definition}(\textbf{Leaf Node})
  \label{def:node2}
   A leaf node is denoted as $N_{l} = (\mathbf{C}_{l}, c)$, where $\mathbf{C}_{l}$ represents the data point set, and $c$ is the leaf node capacity.
\end{definition}

An example of balanced multi-way KD-tree construction is shown in Fig.~\ref{fig:multi-kd}. The construction commences by partitioning the space by selecting two pivots, denoted as $o_1$ and $o_2$, along the specific dimension. One-third of the points lie below $o_1$ along that dimension, while two-thirds lie below $o_2$, ensuring that the dataset is evenly distributed into three sub-spaces. We then recursively select each dimension and repeat the partitioning process until the number of points in each node falls below the leaf node capacity $c=2$. 


\begin{definition}(\textbf{$k$ Nearest Neighbors ($k$NN) Search})
  \label{def:$k$NNQ}
   Given a dataset $\mathbf{D}=\{\mathbf{p}_i\}_{i=1}^{n} \in \mathbb{R}^{n \times d}$, a query point $\mathbf{p}^* \in \mathbb{R}^d$, and an integer $k \in \mathbb{Z}^+$, $k$NN returns $k$ data points in $\mathbf{D}$ that are the most similar to $\mathbf{p}^*$, i.e., it retrieves $\mathbf{P} \subseteq \mathbf{D}$ with $k$ points such that: $\forall \mathbf{p} \in \mathbf{P}$ and $\forall \mathbf{p}' \in \mathbf{D} - \mathbf{P}$, we have $\text{dist}(\mathbf{p}^* ,\mathbf{p}) \leq \text{dist}(\mathbf{p}^*,\mathbf{p}')$. 
\end{definition} 

\begin{definition}(\textbf{Radius Search (RS)})
   \label{def:rangequery}
    Given a dataset $\mathbf{D}=\{\mathbf{p}_i\}_{i=1}^{n} \in \mathbb{R}^{n \times d}$, a query point $\mathbf{p}^* \in \mathbb{R}^d$, and a search radius $r \in \mathbb{R}$, the radius search retrieves all points $\mathbf{p} \in \mathbb{R}^d$ that satisfy the condition $\text{dist}(\mathbf{p},\mathbf{p}^*) \leq r$.
\end{definition}

\subsection{Common Pruning Technologies}
\label{sec:pruning}
We introduce the concepts of \textit{Minimum Bounding Rectangle} and \textit{Minimum Bounding Ball}, along with pruning strategies aimed at avoiding unnecessary computations in queries.

\begin{definition}(\textbf{Minimum Bounding Rectangle (MBR)})\label{def:MBR}
Given a dataset $\mathbf{D}=\{\mathbf{p}_i\}_{i=1}^{n} \in \mathbb{R}^{n \times d}$, a subset $\mathbf{P}\subseteq \mathbf{D}$, the minimum bounding rectangle, denoted by $R =(\blb,\bub)$, is the smallest $d$-dimensional bounding rectangle encompassing all points in $\mathbf{P}$, where $\bub=(tp 
    _{1},tp
    _{2},\cdots,tp_{d}) \in \mathbb{R}^{d}$ is the upper bound, while $\blb=(bp_{1},bp_{2},\cdots,bp_{d}) \in \mathbb{R}^{d}$ is the lower bound.
\end{definition}

\begin{definition}(\textbf{Minimum Bounding Ball (MBB)})
	\label{def:MBB}
Given a dataset $\mathbf{D}=\{\mathbf{p}_i\}_{i=1}^{n} \in \mathbb{R}^{n \times d}$, a subset $\mathbf{P}\subseteq \mathbf{D}$, the MBB, denoted as $B=(\mathbf{o},r)$, is the smallest $d$ dimensional bounding ball containing all the points in $\mathbf{P}$, with its center located at the centroid $\mathbf{o} \in \mathbb{R}^{d}$. The radius of MBB is calculated by $r = \max_{\mathbf{p} \in \mathbf{P}}\text{dist}(\mathbf{p},\mathbf{o})$.
\end{definition}


\begin{lemma}(\textbf{Rectangle-Rectangle Pruning})
  \label{lemma:intersect} 
Given a dataset $\mathbf{D}=\{\mathbf{p}_i\}_{i=1}^{n} \in \mathbb{R}^{n \times d}$, a subset $\mathbf{P}\subseteq \mathbf{D}$, which is compressed by the MBR, $R_1= (\blb, \bub)$, where $\bub=(tp 
    _{1},tp
    _{2},\cdots,tp_{d}) \in \mathbb{R}^{d}$ and $\blb=(bp 
    _{1},bp
    _{2},\cdots,bp_{d}) \in \mathbb{R}^{d}$, and a query range $R_2 = (\blb^{1}, \bub^{1})$, where $\blb^1=(bp 
    _{1}^1,bp
_{2}^1,\cdots,bp_{d}^1) \in \mathbb{R}^{d}$ and  $\bub^1=(up 
    _{1}^1,up
    _{2}^1,\cdots,up_{d}^1) \in \mathbb{R}^{d}$, if the following equation holds:
\begin{equation}\small
\max(bp_i,bp^1_i) \leq \min(tp_i,tp^1_i),\, \forall i=1,\cdots,d,
\end{equation}
then the two MBRs intersect.
\end{lemma}

Since $R_1$ intersects with $R_2$, pruning $R_2$ is not feasible.

\begin{lemma}(\textbf{Ball-Ball Pruning})
\label{lemma:Ball_filter}
Given a dataset $\mathbf{D}=\{\mathbf{p}_i\}_{i=1}^{n} \in \mathbb{R}^{n \times d}$, a subset $\mathbf{P}\subseteq \mathbf{D}$, which is compressed by the MBB, $B=(\bo,r_{o})$ with a center located at the data point $\bo \in \mathbb{R}^{d}$ and the radius $r_o \in \mathbb{R}^+$, a query point $\mathbf{p}^* \in \mathbb{R}^{d}$, and a search radius $r \in \mathbb{R}^+$, if $dist(\bo, \mathbf{p}^*) > r_{o}+r$, the points in $\mathbf{P}$ can be pruned.
\end{lemma}


\begin{lemma}(\textbf{Rectangle-Ball Pruning})
\label{lemma:Rectangle_filter}
Given a dataset $\mathbf{D}=\{\mathbf{p}_i\}_{i=1}^{n} \in \mathbb{R}^{n \times d}$, a subset $\mathbf{P}\subseteq \mathbf{D}$ compressed by the MBR, $R=(\blb,\bub)$, where $\bub=(tp 
    _{1},tp
    _{2},\cdots,tp_{d}) \in \mathbb{R}^{d}$ and $\blb=(bp 
    _{1},bp
    _{2},\cdots,bp_{d}) \in \mathbb{R}^{d}$, and a query point $\mathbf{p}^* = (x_1,x_2,\cdots,x_d) \in \mathbb{R}^{d}$ with a search radius $r$, the minimum distance between $\mathbf{p}^*$ and the MBR, represented by $\text{dist}_{min}(\mathbf{p}^*,R)$, is calculated as following:
\begin{equation}\small 
\label{eqn:MD}
\text{dist}_{min}(\mathbf{p}^*,R) = \sqrt{\sum_{i=1}^{d}(x_i-x')^2}, \, x'=\left\{
\begin{array}{rcl}
bp_{i}, & & {x_i < bp_{i}};\\
tp_{i}, & & {x_i > tp_{i}};\\
x_i, & & {otherwise}.
\end{array} \right. 
\end{equation}

Then $\mathbf{P}$ is pruned if $\text{dist}_{min}(\mathbf{p}^*,R)$ is greater than $r$.
\end{lemma}


 
 \section{Fast Construction of Balanced Index}
\label{sec:Construction}
Before delving into the balanced multi-way KD-tree construction process, we accelerate the space partition procedure by introducing two key steps: 1) instead of experimenting to test which \( t \) yields the highest query efficiency, we optimize the number of comparisons on average in queries to find an appropriate partition number in Section~\ref{sec:partition_time}, and 2) we design a lightweight ML model to approximate the CDF of the dataset for pivot set prediction to avoid sorting in Section~\ref{sec:two_layer_linear_model}. Subsequently, we elaborate on the implementation of the method in the construction process in Section~\ref{sec:index_construction}.

\subsection{Determining Partition Number}
\label{sec:partition_time}
We first introduce a key metric, the average external path length (AEPL) \cite{BereczkyDNR16, NeiningerLS13}, to evaluate the average query performance of the balanced multi-way KD-tree. Then we analyze the impact of $t$ on the AEPL of the index. Based on the analysis, we formulate the choice of $t$ for a given dataset as an optimization problem and approximate the solution.

\begin{definition}(\textbf{Average External Path Length (AEPL)})
   \label{def:AEPL}
   Given a dataset $\mathbf{D}=\{\mathbf{p}_i\}_{i=1}^{n} \in \mathbb{R}^{n \times d}$, the AEPL of a tree-structured index built on $\mathbf{D}$, denoted as $H$, is defined as:
\begin{equation}\small
H = \frac{\sum^n_{i=1}e(\mathbf{p}_i)}{n},
\end{equation}
where $e(\mathbf{p}_i)$ denotes the number of comparisons from the root node to the leaf node storing $\mathbf{p}_i$.
\end{definition}
\begin{figure}
	\centering
\includegraphics[width=.47\textwidth]{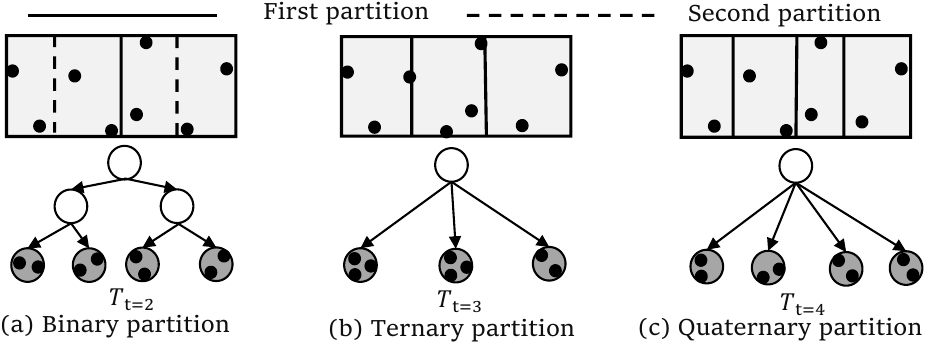}
	\vspace{-0.5em}
	\caption{Impact of \( t \) values on AEPL with leaf node capacity \( c = 3 \). Notably, in (a), since the data points in each subtree after the first partition do not fall below \( c \), a second partitioning is required, represented by the dashed lines.}
	\label{fig:partition}
	\vspace{-1em}
\end{figure}

Based on Definition~\ref{def:AEPL}, for a balanced multi-way KD-tree \( T \), if each leaf node contains the same number of data points \( c_0 \), the AEPL of \( T \), denoted as \( H_T \), can be expressed as:
\begin{equation}\small
\label{eqn:target_problem2}
H_T =[c_0 \times(\frac{t^2}{2}+\frac{t}{2}-1)]^{\lceil log^{\frac{n}{c}}_t\rceil}.
\end{equation}

However, due to the inability to evenly divide an odd number of data points, the distribution results in one leaf node containing one more point than the others, as exemplified by the split of 7 points into groups of 3 and 4. Therefore, the value of \( c_0 \) can be either $\lfloor\frac{n}{t^{(\lceil \log_t \frac{n}{c} \rceil)}}\rfloor$ or $\lceil \frac{n}{t^{(\lceil \log_t \frac{n}{c} \rceil)}}\rceil$.

To approximate \( H_T \), we can determine \( c_0 \) using rounding technology: specifically, we define $\Delta c_0$ as follows:
\begin{equation}\small
   \Delta c_0 =  \frac{n}{t^{\lceil \log_t \frac{n}{c} \rceil}} - \lfloor \frac{n}{t^{\lceil \log_t \frac{n}{c} \rceil}} \rfloor.
\end{equation}

Hence, we acquire \( c_0 \) by comparing \( \Delta c_0 \) with 0.5 using rounding, expressed as follows:
\begin{equation}\small
\label{eqn:second_model}
 c_0 = 
\begin{cases}
\lfloor \frac{n}{t^{\lceil \log_t \frac{n}{c} \rceil}} \rfloor, & \text{if } \Delta c_0 \leq 0.5; \vspace{0.05cm}\\
\lceil \frac{n}{t^{\lceil \log_t \frac{n}{c} \rceil}} \rceil, & \text{if } \Delta c_0 > 0.5;\\
\end{cases}
\end{equation}

\begin{figure}
	\centering
	\vspace{-1.4em}
\includegraphics[width=.47\textwidth]{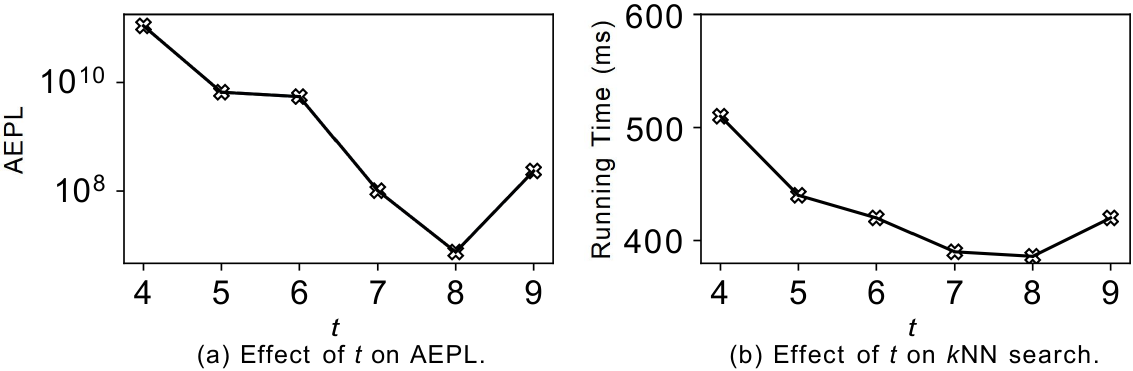}
	\vspace{-0.4em}
	\caption{Impact of different values of $t$ on AEPL and $k$NN efficiency (100 tasks) with leaf capacity $c=30$ on \PC.}
	\label{fig:real_compare}
	\vspace{-1em}
\end{figure}

\myparagraph{Effect of \( t \) on AEPL and Query Efficiency} Adjusting $t$ for a given dataset results in changes in the AEPL value during the construction process. For instance, utilizing $t=2$ as in the balanced KD-tree may not always be the optimal choice to achieve AEPL-optimality across various datasets. In Fig.~\ref{fig:partition}, we implement binary, ternary, and quaternary partitions for index construction, considering $8$ points. The AEPL values for each index are $H_{T_{t=2}} = 2$, $H_{T_{t=3}} = 1.25$, and $H_{T_{t=4}} = 2.25$. Remarkably, $H_{T_{t=3}}$ is the lowest, indicating that $t=3$ achieves AEPL-optimality for constructing the index. Moreover, Fig.~\ref{fig:real_compare} presents the AEPL and $k$NN search efficiency under different $t$ values on 1 million data points from \PC \cite{argoverse}. We observe that as $t$ increases, both AEPL and query runtime (100 query tasks) first decrease and then increase, reaching their minimum at $t=8$. Hence, we use AEPL to measure query efficiency and define the AEPL-optimal criterion as follows:
\begin{definition}(\textbf{AEPL-optimal Criterion})
   \label{def:AEPL-optimal}
   Given a dataset $\mathbf{D} \in \mathbb{R}^{n \times d}$ and a leaf node capacity $c$, 
   a balanced multi-way KD-tree is said to be AEPL-optimal when the following objective is minimized with respect to the partition number $t \in \mathbb{Z}^+$:
   \begin{equation}
   \label{eqn:target_problem}
   H_T = \min_{t \in \mathbb{Z}^+} \left[c_0 \times \left(\frac{t^2}{2} + \frac{t}{2} - 1\right)\right]^{\lceil \log_t^{\,\frac{n}{c}} \rceil}.
   \end{equation}
\end{definition}

A classical way to obtain the minimum of an objective function involves taking its derivative and setting the derivative equal to 0 to find critical points. However, the rounding symbol and $t \in \mathbb{Z}^+$ render $H_T$ non-continuous, leading to a lack of differentiability. To address the limitation, we adopt the simulated annealing approach \cite{SarfiKCKRMB23} to efficiently search for the optimal partition number $t^*$, where candidate values of $t$ are iteratively evaluated and probabilistically accepted to balance exploration and convergence.

\subsection{Accelerating Pivot Set Finding}
\label{sec:two_layer_linear_model}
Once $t^*$ is determined, instead of relying on sorting, which needs to locate each pivot by determining the position of each data point in the dataset, our focus shifts to identifying the positions of pivots. Therefore, we design a lightweight ML model called \textit{two-stage regression model} to avoid unnecessary computations by using a small sample to approximate the CDF of the dataset, then utilizing the CDF to filter out irrelevant points and predict the pivot set in the remaining points.

\myparagraph{Two-stage Regression Model} To estimate the CDF of a dataset in a given dimension $\mathbf{x} = \{x_1, x_2, \cdots, x_{n_1}\}$ and fit arbitrary nonlinear forms, we propose a two-stage regression model. The model first clusters similar data and then applies separate linear models to each cluster, allowing local trends to be captured and improving the overall fitting performance.

The model consists of two stages: the first stage contains the root model, responsible for estimating the CDF value for each data point and clustering data points with similar CDF values into $l$ clusters, and the second stage consists of $l$ sub-models, each trained on the clusters to precisely compute their respective CDFs. We generate samples to support the root model, which approximates the CDF of the dataset and divides the dataset into $l$ clusters based on CDF values, each assigned an ID. The mapping between $x_i$, where $i=1,2,\cdots,n_1$, and the ID of clusters $U = \{u| u<l,\, u \in \mathbb{Z}^+\}$ is expressed as:
\begin{equation}\small
u= l \alpha x_i+l \beta,
\end{equation}
where \( \alpha \) represents the slope, \( \beta \) signifies the intercept, and our objective is to minimize the sum of squared residuals (SSR).

Two notable points are worth mentioning: 1) to ensure differentiability when utilizing derivative-based methods for finding the optimal solution, we treat $u$ as a continuous value, and 2) due to $l \in \mathbb{Z}^+$, obtaining the optimal solution through partial derivatives is not feasible. Instead, we calculate the partial derivatives of the objective with respect to $\alpha$ and $\beta$, and test different values of $l$ to find the optimal solution (see Section~\ref{sec:index}). Hence, we take derivatives for $\alpha$ and $\beta$ and set each derivative to 0, the derived expressions for $\alpha$ and $\beta$ with respect to $l$ can be presented as follows:
\begin{algorithm}[t]\small
	\caption{\texttt{CDF\_Training}($\mathbf{x}'_{n_1}$)}
	\KwIn{$\mathbf{x}'_{n_1}$: the training set.}
    \KwOut{$\mathbf{M}$: the CDF model.}
	\label{alg:CDF_Training}
		$\mathbf{x}_{n_1}' \gets$ Generate sample over $\mathbf{x}$\;\label{line: bb}
        \texttt{Sort}$(\mathbf{x}'_{n_1})$\;  \label{line: quicksort}
	$\mathbf{T}\gets [][]$\tcp*[r]{2-dimension array}\label{line: array1}
        $\mathbf{T}_0 \gets []$\tcp*[r]{1-dimension array}\label{line: array1} \label{line: array2}
	\For{$i \gets 0$ up to $|\mathbf{\mathbf{x}'_{n_1}}|$}{
	    $\mathbf{T}_0$.add(($\mathbf{x}'_{n_1}[i]$, $i/|\mathbf{\mathbf{x}'_{n_1}}|$))\;\label{line: array3}
	}
        $\mathbf{M} \gets []$\tcp*[r]{1-dimension array}
        The root model $\mathbf{M}[0]$ is trained on $\mathbf{T}$[0][0]\;\label{line: a} 
        Partition $\mathbf{T}_0$ into $\mathbf{T}$ by calling $\mathbf{M}[0]$\;
	\For{$i \gets 0$ up to $l$}{ 
	    Training the sub-model $\mathbf{M}[i+1]$ on $\mathbf{T}[i]$\;\label{line: data_set} \label{line: b}
	    }
        \KwRet $\mathbf{M}$\;
\end{algorithm}

\begin{equation}\small
\label{eqn:results}
\begin{pmatrix}
    \alpha \\
    \beta     
\end{pmatrix}
=\frac{1}{l}
\begin{pmatrix}
    \sum^n_{i=1}x_i^2 & \sum^n_{i=1}x_i\\
    \sum^n_{i=1}x_i & \sum^n_{i=1}1    
\end{pmatrix}^{-1}
\begin{pmatrix}
    \sum^n_{i=1}x_i  u_i \\
        \sum^n_{i=1} u_i   
\end{pmatrix}.
\end{equation}

We use symbols to simplify Equation~(\ref{eqn:results}), like $S_x = \sum^n_{i=1}x_i$, $S_{x^2} =\sum^n_{i=1}x_i^2$, $S_u = \sum^n_{i=1}  u_i$, and $S_{xu} = \sum^n_{i=1}x_i u_i$. Hence, $\alpha$ and $\beta$ can be expressed as follows:
\begin{equation}
\begin{split}
\alpha = &\frac{1}{l} \times \frac{n \times S_{xu}-S_{x} \times S_{u}}{n \times S_{x^2}- (S_{x})^{2}}\,,\beta = \frac{1}{l} \times \frac{S_{u}-\alpha \times S_{x}}{n}.
\end{split}
\end{equation}

After segmenting the dataset via the root model, we train each sub-model on the corresponding cluster. Instead of minimizing SSR, employing \textit{piecewise linear fitting} (PLF) \cite{WangMW22} is a preferable choice. This is because, although both models can estimate the CDF effectively, minimizing SSR necessitates traversing all data points to obtain $S_x$, $S_u$, $S_{x^2}$, and $S_{xu}$. In contrast, employing PLF only requires obtaining the maximum and minimum values of the dataset to support model training, thereby reducing the time complexity.

\myparagraph{CDF Model Training}
Algorithm~\ref{alg:CDF_Training} outlines the process for training the CDF model, denoted as $\mathbf{M}$. We use $\mathbf{T}_0$ to represent the training set for the root model (line~\ref{line: array1}) and $\mathbf{T}$ to represent an array storing training clusters for the sub-models (line~\ref{line: array2}). All $(\mathbf{x}'[i], \frac{i}{|\mathbf{x}'|})$ pairs are inserted into $\mathbf{T}_0$, where $\frac{i}{|\mathbf{x}'|}$ represents the CDF value for the training tuple, calculated as its index in the sorted dataset divided by the dataset size $(|\mathbf{x}'|)$ (line~\ref{line: array3}). Each training tuple is processed by the root model, which predicts a CDF value scaled by the number of sub-models. The training tuple is then divided into multiple training subsets, which are used to train the sub-models.
\begin{algorithm}[t]\small
	\caption{\texttt{Build\_Index}($\mathbf{D}$, $d'$)}
        \label{alg:construction}
        \KwIn{$\mathbf{D}$: the dataset, $d'$: the tree depth.}
        \KwOut{$N$: the root node.}
        \textbf{Default Values: }{$i \gets 0$, $j \gets (n-1)$, $d' \gets 0$, Root Node $N$.}
        
        $d' \gets d' \text{ mod } d$\; $\mathbf{X} \gets \{x_{i_1d'}\}_{i_1=i}^{j}$\;
	\If{$|\mathbf{D}| < c$} {
	        Store the vectors in $N$ and \KwRet $N$\;\label{line: nodebreak}
	}
        Determine $t$ using the AEPL-optimal criterion\; \label{line: tdetermine}
        $\mathbf{F} \gets$ \texttt{CDF\_Training($\mathbf{X}$)}\;\label{line: CDF} \label{line: CDF_Estimation}
        Predict the pivot set $\mathbf{O}$ by calling $\mathbf{F}$\;
        Partition $\mathbf{D}$ into $\mathbf{D}_1, \mathbf{D}_2, \cdots, \mathbf{D}_{(t-1)}$ using $\mathbf{O}$ and $d'$\;\label{line: partition}
        \For{$i \gets 0$ up to $(t-1)$}{ 
	    $N.\text{child}[i] \gets$ \texttt{Build\_Index}($\mathbf{D}_i$, $d'$)\;
	}
\end{algorithm}
\begin{figure}
\vspace{-1em}
	\centering  \includegraphics[width=.47\textwidth]{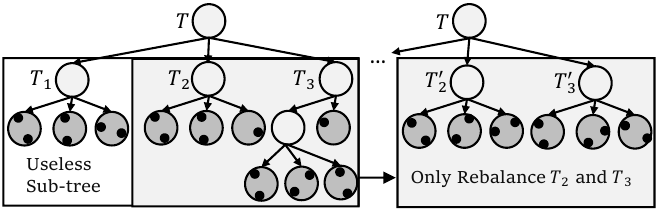}
	\vspace{-0.4em}
	\caption{An example of the rebalance process.}
	\label{fig:rebalance}
	\vspace{-1em}
\end{figure}
\subsection{Time-efficient Construction Process} 
\label{sec:index_construction}
After introducing the above two key steps to accelerate space partitioning, we explore their implementation in the construction process. Algorithm~\ref{alg:construction} shows the construction process, and we first calculate $t$ (see Section~\ref{sec:partition_time}) and estimate the CDF model $\mathbf{M}$ by the two-stage regression model (line~\ref{line: CDF_Estimation}). Once $\mathbf{M}$ is obtained, we use it to exclude points that cannot be the pivot set $\mathbf{O}$ and find $\mathbf{O}$ among the remaining candidates.

Specifically, each pivot $o \in \mathbf{O}$ has a corresponding quantile $q$, such as the median's quantile being $0.5$. To predict each pivot $o$, we utilize $\mathbf{M}$ to scan the dataset, including the data in the pivot candidate set when the quantile falls within the range $[q-\kappa, q+\kappa]$, where $\kappa$ represents a predefined error bound and lies within the range $[0, \frac{1}{t}]$. Taking the prediction of the median as an example, $\kappa$ lies within the range $[0, 0.5]$. If $\kappa = 0$, then the interval $[0.5,0.5]$ solely contains the value corresponding to $0.5$, potentially leading to an empty pivot candidate set. However, if $\kappa = 0.5$, the prediction range expands to $[0, 1]$, including all data points as pivot candidates. Section~\ref{sec:index} suggests that 0.15 is a suitable choice.  

Further, for each pivot $o$, we collect pivot candidates falling into $[q-\kappa, q+\kappa]$, sort them, and identify the median as $o$. Finally, $\mathbf{D}$ undergoes partitioning into subsets based on the predicted pivots (line~\ref{line: partition}), and we continue this process recursively until $\mathbf{D}$ reaches the leaf node capacity $c$ (line~\ref{line: nodebreak}).

\myparagraph{Time Complexity Analysis} The time complexity equals the product of the time complexity of partition in each level and tree depth $h$ ($h<<n$). At each level, the dataset is divided into $t$ subsets, with each subset having $n_i$ data points, and $\sum^{t}_{i=1}n_i = n$. For each subset, the time complexity comprises three parts: 1) the training process involves sampling and sorting, costing $O(\delta n_i+\delta n_i  \log(\delta n_i))$, where $\delta$ is the sampling rate; 2) using the two-stage regression model to train $\mathbf{M}$ costs $O(5\delta n_i)$; 3) predicting the pivot set and partitioning the subset cost $O(2 \delta n_i)$. Hence, the time complexity for each subset is
$
O(8h\delta n_i + h\delta n_i \log(h\delta n_i)),
$
where $\mathbf{n} = (n_1, n_2, \dots, n_t)$ and $\sum_{i=1}^{t} n_i = n$.
To derive the time complexity, we maximize this expression with respect to $\mathbf{n}$, as shown below:
\begin{equation}\small
\begin{split}
    &\sum^{t}_{i=1}(8h\delta n_i+h\delta n_i \log(h\delta n_i))  \leq  8h\delta n+h\delta n \log(h \delta n),
\end{split}
\end{equation}
where $h<<n$ and thus we can regard $h$ as a constant. Considering the highest order term in time complexity analysis, the time complexity is \( O(h\delta n \log(h \delta n)) \).

\section{Real-time In-place Insertion}
\label{sec:Insertion}
Before discussing the in-place insertion, we propose the selective sub-tree rebuilding scheme to improve its efficiency by expediting the rebalancing process. The scheme improves rebalance efficiency by excluding unnecessary parts of the unbalanced sub-tree rather than opting for an entire rebuild.

\subsection{Selective Sub-tree Rebuilding Scheme} 
\label{sec:Rebuilding_Strategy1}
We find that the scapegoat strategy \cite{yixi2021} may have unnecessary computations during the full rebuilding of unbalanced sub-trees, resulting in additional time costs. For example, in Fig.~\ref{fig:rebalance}, the sub-tree $T$ is identified as unbalanced, necessitating reconstruction involving 18 data points. However, to rebalance $T$, it is only necessary to reconstruct $T_2$ and $T_3$, involving 12 data points. To accelerate the rebalancing process, we propose a selective sub-tree rebuilding scheme to filter out the irrelevant data points in reconstructions. To be specific, we first explore how to determine whether the tree is unbalanced based on the \textit{$\omega$-balanced criterion} \cite{yixi2021}. Next, we identify which parts of the unbalanced sub-trees do not need reconstruction by dynamically maintaining two variables.

\begin{definition}(\textbf{$\omega$-balanced Criterion})
   \label{def:balancedCriterion}
    Considering a sub-tree rooted at node $N$ within the incremental index, the sub-tree is $\omega$-balanced if and only if it meets the following condition:
\begin{equation}\small
\label{eqn:balancedCriterion}
\mathcal{S}(T(N.\mathbf{C}_{nl}[i])) <\frac{1}{(t-1)} \times \omega \times \mathcal{S}(N),
\end{equation}
where $i =1,2,\cdots,t $, $\omega \in (0.5,1)$ and $\mathcal{S}(N)$ is the number of data points in the sub-tree rooted at $N$. 
\end{definition} 

We can extend Definition~\ref{def:balancedCriterion} to a set of sub-trees. Specifically, a set of sub-trees rooted at \( N \), denoted as \( T(\{N.\mathbf{C}_{nl}[j]\}_{j=i_0}^{j=i_1}) \), where \( 0 < i_0 < i_1 < t \), represents the sub-trees from \( i_0 \)-th to \( i_1 \)-th of $N$. If it satisfies the following inequality, \( T(\{N.\mathbf{C}_{nl}[j]\}_{j=i_0}^{i_1}) \) is \(\omega\)-balanced:
\begin{equation}\small
\label{eqn:balancedCriterion2}
\mathcal{S}(T({{N.\mathbf{C}_{nl}[j]\}}^{j=i_1}_{j=i_0}})) <\frac{i_1-i_0}{(t-1)} \times \omega \times \mathcal{S}(N).
\end{equation}
If \( T(N.\mathbf{C}_{nl}[i]) \), where \( 0 < i < t \), is unbalanced, we do not need to rebalance \( T(N) \); Otherwise, we define the sub-trees needing reconstruction as \( T_{reb} \) and add \( T(N.\mathbf{C}_{nl}[i]) \). We then recursively incorporate neighboring sub-trees of \( T(N.\mathbf{C}_{nl}[i]) \) until Inequality~\eqref{eqn:balancedCriterion2} is satisfied, and finally, reconstruct \( T_{reb} \).

\myparagraph{Pruning Unnecessary Rebalance Parts} The sub-tree \( T(N.\mathbf{C}_{nl}[i]) \) is unbalanced. We dynamically maintain two variables to minimize the number of data points in the sub-tree for rebuilding. The two variables are denoted as $\zeta$, representing the number of data points in the sub-trees, and \( \psi \), defined as a tuple \( (i_0, i_1) \), where \( i_0 \leq i \leq  i_1 \), indicating the sub-trees rooted at the children of \( N \) from the \( i_0 \)-th to the \( i_1 \)-th. Starting from the $i$-th sub-tree, we add nearby sub-trees and check if they satisfy the $\omega$-balance criterion. This involves iterating under all conditions $(i_1, i_0)$, where $0 < i_0 < i$ and $0 < i_1 < t$, stopping the iteration when $T({{N.\mathbf{C}_{nl}[j]\}}^{j=i_1}_{j=i_0}})$ is balanced. Each time a new sub-tree is added, we update $\zeta = \mathcal{S}(T({{N.\mathbf{C}_{nl}[j]\}}^{j=i_1}_{j=i_0}}))$ and $\psi = (i_0, i_1)$ if the following condition is satisfied:
\begin{equation}
    \mathcal{S}(T({{N.\mathbf{C}{nl}[j]}}^{j=i_1}{j=i_0})) > \zeta.
\end{equation}

Therefore, the corresponding $b$ indicates the sub-trees rooted at the children of $N$ that require reconstruction.

\subsection{Insertion with Dynamic Rebalancing}
After discussing the rebuilding scheme, we delve into its implementation in the insertion process, as outlined in Algorithm~\ref{alg:insertion}. The algorithm involves a backtracking approach for insertion. In the process of each backtrack, we examine whether $N$ violates the $\omega$-balance criterion and necessitates rebuilding using the selective sub-tree rebuilding scheme (lines~\ref{line: judge_rebuild1}-\ref{line: judge_rebuild2}) to identify which part needs to be rebuilt. Once identified, instead of using the entire dataset to retrain the two-stage regression model (see Section~\ref{sec:two_layer_linear_model}), we only use the changed dataset to update the model. For newly inserted data points, we denote them as $S^1_x = \sum^{n'_1}_{i=1} x_i$, and for points allocated to other sub-trees, we represent them as $S^2_x = \sum^{n'_2}_{i=1} x_i$. Hence, $S'_{x}$ for updating can be expressed as:
\begin{equation}\small
S'_{x} = S_{x}+S^1_{x}-S^2_{x}.
\end{equation}

Similarity, we can define the $S'_{u}$, $S'_{xu}$ and $S'_{x^2}$. Hence, updating the root model parameter $\alpha$ is expressed as follows:
\begin{equation}\small
\alpha = \frac{(n'+n'_1-n'_2) \times S'_{xu}-S'_{x} \times S'_{u}}{(n'+n'_1-n'_2) \times S'_{x^2}- (S'_{x})^{2}}, 
\end{equation}
and then the expressions for updating $\beta$ is as follows:
\begin{equation}\small
\beta = \frac{S'_{u}-\alpha \times S'_{x}}{(n'+n'_1-n'_2)}.
\end{equation}

Therefore, we update the root model and allocate data points to each sub-model for training, thereby updating the two-stage regression model and making predictions for the pivot set.
\begin{algorithm}[t]\small
	\caption{\texttt{In\_Place\_Insertion}($\mathbf{D}_1$, $N$, $d'$)}
   \KwIn{$\mathbf{D}_1 \in \mathbb{R}^{n_1 \times d}$: the new coming data, $N$: the node, $d'$ is the tree depth.}
   \label{alg:insertion}
   \textbf{Default values: }$N \gets$ the root node, $d' \gets 0$.\
   
        \If{$\mathbf{D}_1$ is null}{ 
             \KwRet \;  
        }
        \If{\texttt{$N$ breaks the $w$-balanced-criterion}}{ \label{line: judge_rebuild1}
             Rebuild the sub-tree rooted at $N$\;  \label{line: judge_rebuild2}
        }
        \eIf{$N.isleaf$}{ \label{line: leafnode}
	       $N$.add($\mathbf{D}_1$)\; \label{line: inserted}
           \If{The number of vectors in $N$ exceeds $c$}{
	       Split $N$ into new leaf nodes\; \label{line: rebuild}
        }  
        }{
            $d' \gets d' \text{ mod } d$\;
            $\mathbf{D^*} \gets [][]$ \tcp*[r]{2-dimension array}
            \For{$i \gets 0$ up to $|N.{\mathbf{O}}|$}{\label{line: bulkloading1}
                 \For{$j \gets 0$ up to $n_1$}{
                     \If{$N.{\mathbf{O}}[i][d']<\mathbf{p_j}[d'] < N.{\mathbf{O}}[i][d'] $}{
                         $\mathbf{D^*}[i].\text{add}(\mathbf{p_j})$\;
                     }
                 }
            }\label{line: bulkloading2}
            \ForEach{$i \gets 0$ up to $|N.{\mathbf{O}}|$}{
                 \texttt{In\_Place\_Insertion}($\mathbf{D^*}[i]$, $N$, $d'$+1);\label{line: bulkloading3}
            }            
        }
\end{algorithm}
\begin{figure}
	\centering
 \vspace{-0.5em}
  \includegraphics[width=.47\textwidth]{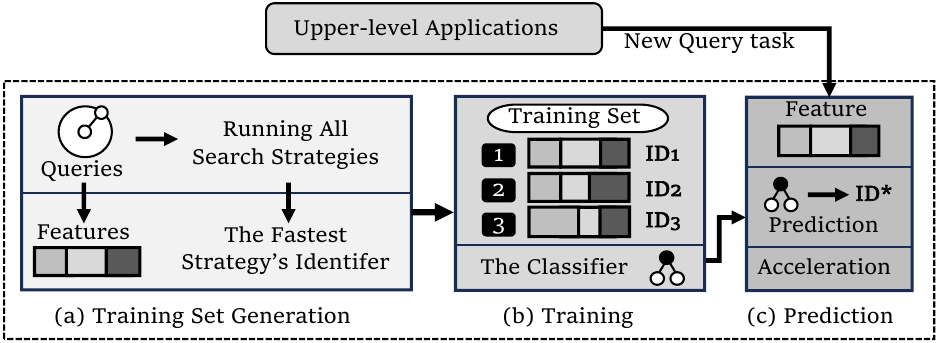}
	\vspace{-0.5em}
	\caption{Three modules in our auto-selection model.}
	\label{fig:autoselection1}
	\vspace{-1em}
\end{figure}
For inserting data points, repeatedly invoking the same function requires additional operations, such as restoring the execution state from the call stack, thereby increasing time costs. To mitigate the frequent invocation of the insertion function, we bulk-load data points (lines~\ref{line: bulkloading1}-\ref{line: bulkloading2}). Specifically, we group the points belonging to a specific sub-space into a set and insert them into the corresponding sub-tree (line~\ref{line: bulkloading3}).

\section{Search Strategy Auto-selection}
\label{sec:auto}
We design an auto-selection model to improve queries' efficiency by automatically selecting the optimal search strategy for the arbitrary query task. We initially present the search strategies for conducting $k$NN and radius searches over \learn. Then, as shown in Fig.~\ref{fig:autoselection1}, we extract query meta-features from search tasks and generate class labels (i.e., ground truth) from the training logs, containing the best search strategy for each query. Next, we feed the query meta-features and class labels into a classification model to learn how to map from a query task to the best-performing strategy. Using the trained model, we predict the optimal strategy for new queries.

\begin{algorithm}[t]\small
\caption{\texttt{UnIS}\_\texttt{$k$NN}($\bq$, $k$, $B$, $T$)}
\KwIn{$\bq$: the query point, $k$: an integer, $B$: the bounding volume, $T$: the traversal strategy.}
\KwOut{$Q$: the priority queue.}
\label{alg:kNN_query}

$Q \gets$ empty priority queue of size $k$\;
$N \gets$ root node\;

\eIf{$T$ is the DFS}{
   \textsf{DFS}($\bq$, $N$, $B$, $k$, $Q$)\;\label{line:dfs_call}
}{
   \textsf{BFS}($\bq$, $N$, $B$, $k$, $Q$)\;\label{line:bfs_call}
}
\KwRet $Q$\;

\vspace{-0.5em}
\algrule
\vspace{-0.5em}

\SetKwFunction{FDFS}{\textsf{DFS}\!}
\SetKwProg{Fn}{Function}{}{}
\Fn{\FDFS{$\bq$, $N$, $B$, $k$, $Q$}}{
   \If{$N$ is a leaf}{\label{line:dfs_leaf_start}
      \ForEach{point $\bp$ in $N$}{
         $ED \gets \text{dist}(\bq, \bp)$\;
         $Q$.add(($\bp$, $ED$))\;
      }\label{line:dfs_leaf_end}
   }
   \tcc{Lemma \ref{lemma:intersect} is used.}
   \If{pruning condition based on $B$ is not met or $Q.size() < k$}{\label{line:dfs_prune_start}
      Recurse to child nodes\;\label{line:dfs_prune_end}
   }
}

\vspace{-0.5em}
\algrule
\vspace{-0.5em}

\SetKwFunction{FBFS}{\textsf{BFS}\!}
\SetKwProg{Fn}{Function}{}{}
\Fn{\FBFS{$\bq$, $N$, $B$, $k$, $Q$}}{
   $Q_1 \gets$ queue initialized with $N$\;
   \While{$Q_1$ is not empty}{
      $N \gets Q_1$.pop()\;
      \If{$N$ is a leaf}{\label{line:bfs_leaf_start}
         \ForEach{$\bp$ in $N$}{
            $ED \gets \text{dist}(\bq, \bp)$\;
            $Q$.add(($\bp$, $ED$))\;
         }\label{line:bfs_leaf_end}
      }
      \Else{
        \tcc{$B=$ MBR uses Lemma \ref{lemma:Ball_filter}, and $B=$ MBB uses Lemma \ref{lemma:Rectangle_filter}.}
         \If{pruning condition based on $B$ is not met or $Q.size() < k$}{\label{line:bfs_prune_start}
            Enqueue all children of $N$ into $Q_1$\;\label{line:bfs_prune_end}
         }
      }
   }
}
\end{algorithm}

\subsection{Search Strategies}
\label{sec:Search_Strategies}
Numerous factors could impact the query efficiency, and considering all possible factors in the search strategies is computationally impractical. Therefore, we narrow our focus to two key factors: traversal strategies (\dfs or \bfs) and the minimum bounding volume (\ball or \rectangle), resulting in four search strategies. Table~\ref{tab:pruning} outlines the pruning strategies corresponding to each search strategy for $k$NN and radius search. For example, $R_{\dfs}$ represents a search strategy employing the \dfs and the MBR. Specifically, in $k$NN search, Lemma~\ref{lemma:Rectangle_filter} is utilized to prune search spaces, while in radius search, Lemma~\ref{lemma:intersect} is employed for the same purpose. Due to space limitations, we only provide the $k$NN procedure here, while the detailed radius search process is included in the appendix (see Section~\ref{sec:rs}). In $k$NN, both traversal strategies and pruning are essential for query efficiency, whereas in radius search, only traversal strategies have a notable impact.
\begin{figure}
	\centering
 \vspace{-1.4em}
\includegraphics[width=.47\textwidth]{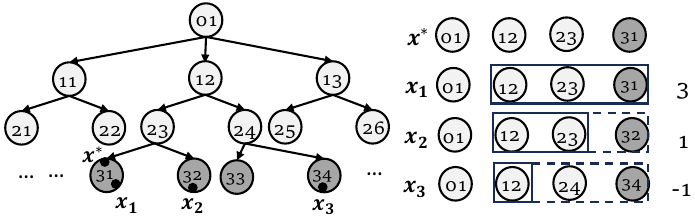}
	\vspace{-1.2em}
	\caption{An example of the index-based metric.}
	\label{fig:similarity}
	\vspace{-1.5em}
\end{figure}

Algorithm~\ref{alg:kNN_query} outlines the procedure for processing $k$NN queries. Depending on the traversal strategy, either DFS (line~\ref{line:dfs_call}) or BFS (line~\ref{line:bfs_call}) is used. In DFS, the algorithm recursively explores nodes starting from the root and processes all leaf nodes (lines~\ref{line:dfs_leaf_start}–\ref{line:dfs_leaf_end}). During traversal, pruning is performed using Lemma \ref{lemma:intersect}, regardless of whether $B$ is MBR or MBB. If the pruning condition is not satisfied, the algorithm recurses to the child nodes (lines~\ref{line:dfs_prune_start}–\ref{line:dfs_prune_end}). In BFS, the algorithm processes nodes level by level, starting from the root and proceeding to its children. During the search, Lemma \ref{lemma:Ball_filter} is used when $B$ is the MBR, and Lemma \ref{lemma:Rectangle_filter} is applied when $B$ is the MBB.



\subsection{Ground Truth Generation}
\label{sec:groundtruth} 

\subsubsection{Class Label Generation}
\label{sec:Label_Generation}
We generate queries and process queries using different search strategies. For each query, we extract query meta-features and record the search strategy with the shortest execution time. Notably, we improve sample generation efficiency by terminating strategies that exceed the current best execution time $u$; otherwise, $u$ is updated.
\begin{algorithm}[t]
\caption{\texttt{Auto\_Selection}($\mathbf{D}$, $\mathcal{G}$)}
\KwIn{$\mathbf{D}$: the dataset; $\mathcal{G}$: the query task set.}
\KwOut{$\mathcal{Z}$: the set of predicted search strategies.}
\label{alg:auto_selection}

$\mathcal{G}_0 \gets$ randomly generate a subset of query tasks\; \label{line:auto_sample}
$S \gets$ an empty training set, 
$Z \gets$ an empty strategy set\;

\For{each query task $g \in \mathcal{G}_0$}{ \label{line:auto_training_sample1}
    Extract features $\mathbf{F}$ from $g$\;
    $z^* \gets$ the fastest among the four search strategies\;
    $S.\text{add}(\langle \mathbf{F}, z^* \rangle)$\;
}\label{line:auto_training_sample2}

Train the classifier on $S$\; \label{line:auto_set_training}

\For{each query task $g \in \mathcal{G}$}{\label{line:auto_prediction1}
    Extract features $\mathbf{F}$ from $g$\; 
    Predict the optimal strategy $z^*$ for $g$ based on $\mathbf{F}$\; \label{line:auto_prediction2}
    $\mathcal{Z}.\text{add}(z^*)$\;
}

\KwRet $\mathcal{Z}$\;

\end{algorithm}

\begin{table}[t]
\vspace{-1.5em}
\setlength{\tabcolsep}{7.6pt}
\caption{Pruning technologies on each search strategy.}
\label{tab:pruning}
\begin{tabular}{ccccc}
\toprule
           \textbf{Pruning} & $R_{\dfs}$ & $R_{\bfs}$ & $B_{\dfs}$ & $B_{\bfs}$ \\ 
           \midrule
$k$NN &   Lemma \ref{lemma:intersect}     &      Lemma \ref{lemma:Ball_filter}                &      Lemma \ref{lemma:intersect}                &       Lemma \ref{lemma:Rectangle_filter}               \\
Radius Search &    Lemma \ref{lemma:Ball_filter}                  &          Lemma \ref{lemma:intersect}             &        Lemma \ref{lemma:Ball_filter}              &        Lemma \ref{lemma:intersect}               \\ 
\bottomrule
\end{tabular}
\vspace{-2em}
\end{table}

\subsubsection{Query Meta-feature Extraction}
\label{sec:Graph_feature}
Utilizing each value in the data point $\mathbf{p}^* $ as a query meta-feature, denoted as $\mathbf{F}_1$, is not informative enough. This limitation is evident in the auto-selection model utilized in the \aitree \cite{Abdullah2022}, as depending solely on $\mathbf{F}_1$ does not result in high prediction accuracy. Hence, the sub-optimal strategy predicted by the model with low accuracy leads to an increase in query time.

Instead of only using $\mathbf{F}_1$ as the query meta-feature, we explore the relationships between each data point in the training sample $\mathbf{X}$ to uncover the latent information $\mathbf{F}_2$. Precisely, we learn graphical representations of $\mathbf{X}$ by analyzing the similarity between points based on the index-based metric and building the undirected weighted graph. Derived from the positions of data points in the graph, we extract the graph-based feature of each query point, labeled as $\mathbf{F}_2$. Then, we merge $\mathbf{F}_1$ and $\mathbf{F}_2$ into $\mathbf{F}$ to describe a query task.

\myparagraph{Graph Generation} We generate an undirected weighted graph $G = (\mathbf{X},\mathbf{E})$ where each query point can be represented by a vertex $\mathbf{x}_i \in \mathbf{X}$. Vertices $\mathbf{x}_i$ and $\mathbf{x}_j$ are linked by an edge $e = (\mathbf{x}_i, \mathbf{x}_j) \in \mathbf{E}$, denoting the similarity between two instances, computed using the metric discussed below. 

\myparagraph{Index-based Metric} 
During insertion, each data point (or instance) is compared with the pivots and recursively assigned to sub-spaces. The more frequently points are allocated to the same sub-spaces, the more similar they are. Conversely, the more times they are distributed to different sub-spaces, the more dissimilar they are. Hence, we propose the index-based metric to measure data point similarity:
\begin{definition}(\textbf{Index-based Metric})
For each pair of instances $\mathbf{x}_i$ and $\mathbf{x}_j$ (where $i \neq j$), the index-based metric $\mathcal{I}(\mathbf{x}_i,\mathbf{x}_j)$ is calculated as the times they fall into the same sub-space minus the times they fall into different sub-spaces before reaching the leaf node. The expression is as follows:
\begin{equation}\small
\mathcal{I}(\mathbf{x}_i,\mathbf{x}_j) = \mathcal{T}^+(\mathcal{U}(\mathbf{x}_j),\mathcal{U}(\mathbf{x}_i)) -\mathcal{T}^-(\mathcal{U}(\mathbf{x}_j),\mathcal{U}(\mathbf{x}_i)) -1,
\end{equation}
where $\mathcal{U}(\mathbf{x}_i)$ denotes the node path from the root to the leaf node containing $\mathbf{x}_i$, $\mathcal{T}^+(\mathcal{U}(\mathbf{x}_i),\mathcal{U}(\mathbf{x}_j))$ represents the count of common nodes in the two paths, and $\mathcal{T}^-(\mathcal{U}(\mathbf{x}_i),\mathcal{U}(\mathbf{x}_j))$ signifies the count of different nodes in the two paths.
\end{definition}

In Fig.~\ref{fig:similarity}, we apply $t=3$ to construct the index and label each sub-space in each partition. For example, during the first partition process, sub-spaces are labeled as ``11'', ``12'', and ``13'', representing the three sub-spaces (also known as sub-trees). Then, we compute similarities between $\mathbf{x}^*$ and each of $\mathbf{x}_1$, $\mathbf{x}_2$, and $\mathbf{x}_3$. Each instance undergoes three partitions to reach a leaf node, and we count the times they fall into the same sub-space and different sub-spaces. For example, $\mathbf{x}^*$ and $\mathbf{x}_2$ consistently fall into the same sub-space, such as ``12'' and ``24'', and into different sub-spaces, with $\mathbf{x}^*$ traversing ``31'' and $\mathbf{x}_2$ traversing ``32'', resulting in $\mathcal{I} = (\mathbf{x}^*, \mathbf{x}_1) = 1$.

\myparagraph{Feature Generation} Once the graph is constructed, the process of generating feature $\mathbf{F}_2$ unfolds as follows: Initially, an instance $\mathbf{x}$ enters its corresponding leaf node. Then, edges between $\mathbf{x}$ and other instances can be directly obtained based on the leaf node's position in the index. For example, in Fig.~\ref{fig:similarity}, if an instance resides in the ``34'' leaf node, the edge with all data points in the ``31'' leaf node is 1. Hence, to derive $\mathbf{F}_2$, we traverse all leaf nodes and obtain a fixed-sized feature. 

\subsection{Model Training and Prediction}
\label{sec: model_training}
The process outlined in Algorithm~\ref{alg:auto_selection} involves training the auto-selection model to predict the optimal search strategy, treating the strategy selection as a multi-class classification problem. Initially, we generate a set of query tasks $\mathcal{G}_0$ randomly (line~\ref{line:auto_sample}). Each task undergoes query meta-feature extraction and execution using all available search strategies, with the fastest one identified as the optimal strategy $z^*$ (lines~\ref{line:auto_training_sample1}–\ref{line:auto_training_sample2}).
Next, using the ground truth, we train the multi-class classifier (line~\ref{line:auto_set_training}). Specifically, we use \textit{ensemble learning} (EL) \cite{ZhaoWHYZPS23} (Random Forest \cite{SuZ06} in the experiments), which combines the outputs of multiple learners to address the potential inaccuracies of individual classifiers. Once the auto-selection model is trained, it can predict the optimal search strategy for query processing (lines~\ref{line:auto_prediction1}–\ref{line:auto_prediction2}).
\section{Experiments}
\label{sec:exp}
\begin{table}[]
\setlength{\tabcolsep}{1.1pt}
\ra{1.2}
\caption{An overview of datasets (M for million).}
\label{tab:datasets}
\vspace{-1em}
\scalebox{0.815}{
\begin{tabular}{ccccccccc}
\toprule 
\textbf{Dataset}
& \POI        
& \AVL
& \Porto        
& \Tdrive 
& \Shapenet 
&  \PC
& \Apollo  
& \Trajectory 

\\ \midrule

\textbf{Ref.}
& \cite{argoverse}
& \cite{argoverse} 
& \cite{Proto} 
&  \cite{yuan2010t-drive} 
& \cite{shapenet}
& \cite{argoverse} 
& \cite{Apolloscope} 
& \cite{argoverse} \\

$d$
& 2   
& 2     
& 2
& 2
& 3           
& 3              
& 3    
& 4     \\

\textbf{Scale}
& 6M        
& 2M 
& 1.27M   
& 1.27M 
& 1M
& 10.00M      
& 10.00M  
& 2.70M     \\ \bottomrule
\end{tabular}}
\vspace{-2em}
\end{table}
\begin{table*}[] 
\setlength{\tabcolsep}{4.2pt} 
\caption{Impact of parameter settings on \UnIS construction evaluated on the \POI dataset.} 
\vspace{-0.5em}
\label{tab:small_parameter_learning_normalized} 
\begin{tabular}{cccccccccccccccc} 
\toprule 
\multirow{2}{*}{\textbf{Parameters}} & \multicolumn{7}{c}{Sampling Rate $\delta$} & \textbf{} & \multicolumn{7}{c}{\# Models $l$} \\ 
\cmidrule{2-8} \cmidrule{10-16} 
& $5\times10^{-4}$ & $1 \times 10^{-3}$ & $5\times 10^{-3}$ & $1\times 10^{-2}$ & $5\times 10^{-2}$ & $1 \times 10^{-1}$ & $5 \times 10^{-1}$ & & 10 & 50 & 100 & 500 & 1000 & 5000 & 10000 \\ 
\toprule 

$\frac{t^0}{t^1}$ 
& 0.55 & 0.39 & 0.32 & 0.24 & 0.26 & 0.29 & 0.80 
& & 0.92 & 0.86 & 0.83 & 0.83 & 0.78 & 0.77 & 0.84 \\ 

\vspace{0.5em}
$t_{k\text{NN}}^0/t_{k\text{NN}}^1$
& 1.00 & 0.83 & 0.17 & 0.17 & 0.00 & 0.00 & 0.50 
& & 1.87 & 1.52 & 1.81 & 1.34 & 1.30 & 1.55 & 1.57 \\ 

$t_{\text{RS}}^0/t_{\text{RS}}^0$
& 45.25 & 0.25 & 24.25 & 7.50 & 3.75 & 0.25 & 16.00
& & 2.72 & 2.49 & 2.53 & 2.11 & 1.99 & 2.23 & 1.97 \\ 

\bottomrule
\end{tabular} 
\vspace{-0.5em}
\end{table*}


\subsection{Experimental Setup}
\label{sec:setup}
\myparagraph{Dataset} We evaluate our method on eight real-world datasets: \POI, \AVL, \Porto, \Shapenet, \Tdrive, \PC, \Apollo, and \Trajectory. All datasets were collected on edge devices; for example, \POI, \AVL, \PC, and \Trajectory were obtained from sensors mounted on vehicles, all from an autonomous driving dataset called Argoverse \cite{argoverse}. Table~\ref{tab:datasets} summarizes the details.

\myparagraph{Implementations}
We implement \UnIS and all comparisons using Java 8 and evaluate performance on two platforms. The first platform is a server equipped with an i9-14900KF CPU and 128 GB RAM. In our experiments, 8 CPU cores are used. The server supports parameter learning and large-scale experiments, including indexing, insertion, and query performance. The second platform is a wheeled mobile robot, Wheeltec R550, which serves as an edge device to validate the superior performance of \learn. It is equipped with a Jetson Nano \cite{Nano}, which integrates a 64-bit ARM Cortex-A57@1.43 GHz and 4 GB of memory. The source code is available at \cite{UnIS} for reproducibility. Unless otherwise specified, we set $\delta = 0.01$ and $l = 100$ (see Section \ref{sec:index} for details) for index construction, and our algorithm is single-threaded.

\myparagraph{Comparisons} We first compare our method with existing indexing structures, including \Ball \cite{omohundro1989}, \BDLtree \cite{BDLtree2021}, \kdtree \cite{BereczkyDNR16}, and \aitree \cite{Abdullah2022}. For search strategy auto-selection, we consider several SOTA models as competitors, including \Unik \cite{Wang2020}, the auto-selection model in the ``AI+R''-tree (\aitree) \cite{Abdullah2022}, and \TRIO \cite{Shapira2021}. All comparisons are discussed in Section~\ref{sec:related_sec}.
\begin{figure*}
	\centering \includegraphics[width=1\textwidth]{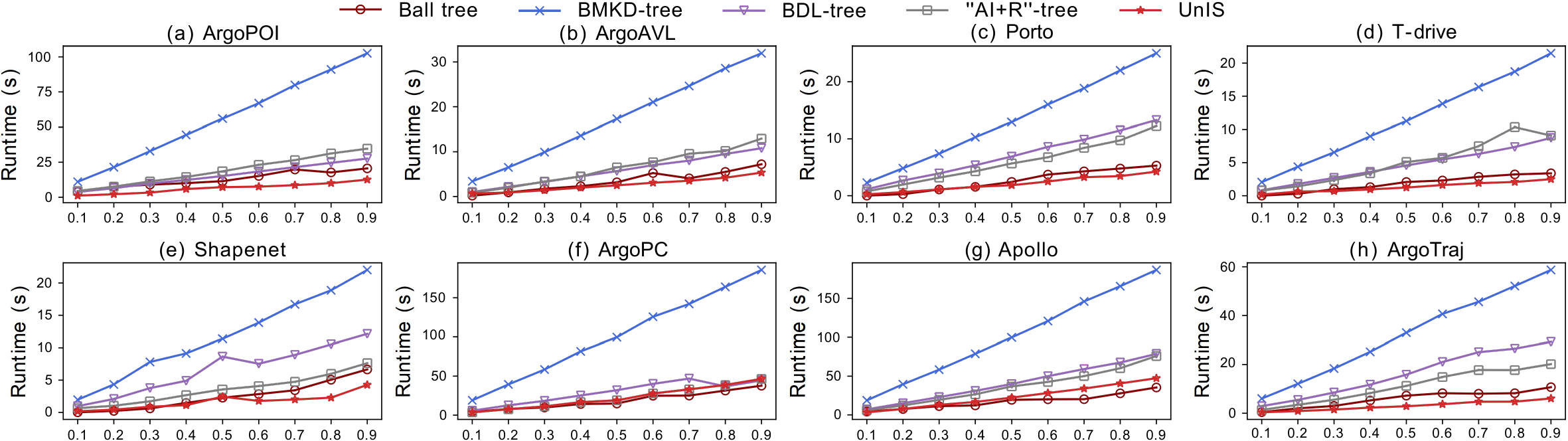}
	\vspace{-1.2em}
	\caption{Comparison of indexing time across different indexes.}
	\label{fig:indextime}
	\vspace{-1.4em}
\end{figure*}

\begin{figure*}
	\centering \includegraphics[width=1\textwidth]{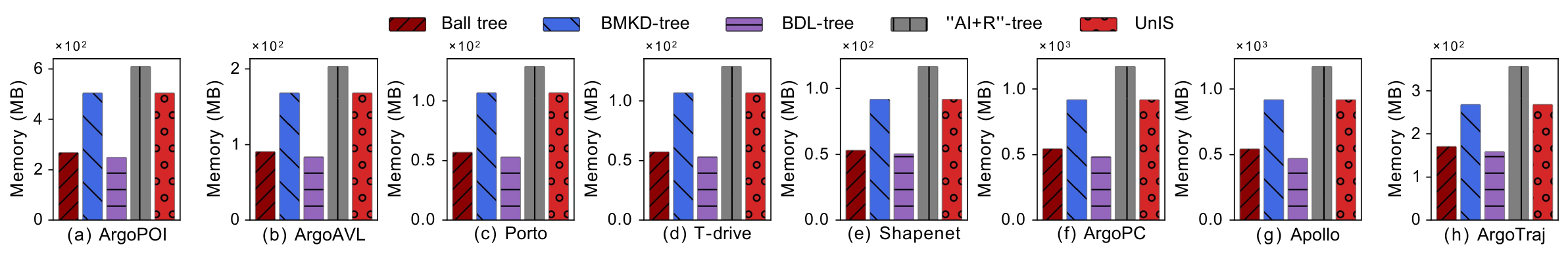}
	\vspace{-2em}
	\caption{The memory cost of each algorithm.}
	\label{fig:memory1}
	\vspace{-1em}
\end{figure*}
\begin{table}[]
\setlength{\tabcolsep}{0.9pt}
\ra{1.2}
\caption{The impact of two factors on $k$NN search.}
\label{tab:BDMM}
\vspace{-1em}
\scalebox{0.82}{
\begin{tabular}{ccccccccc}
\toprule 
\textbf{Dataset}
& \POI        
& \AVL
& \Porto        
& \Shapenet    
& \Tdrive  
&  \PC
& \Apollo  
& \Trajectory 
\\ \midrule
$\frac{t_{\text{BFS}}}{t_{\text{DFS}}}$   & 0.01 & 1.12 & 0.01 & 0.01 & 9.64 & 0.01 & 6.44 & 0.04 \\
$\frac{t_{\text{MBR}}}{t_{\text{MBB}}}$   & 4.34 & 0.22 & 3.04 & 0.04 & 0.06 & 0.15 & 2.40 & 0.03 \\ 
\bottomrule
\end{tabular}}
\vspace{-1.4em}
\end{table}

\subsection{Index Construction}
\label{sec:index}
Parameter studies are conducted on the server due to its large computational capacity. We also use the Wheeltec R550 as an edge device to demonstrate the construction efficiency of \learn\ compared with SOTAs. Note that we evaluate the performance of the two-stage linear model; the detailed results are provided in the appendix (see Section~\ref{sec:two}).

\myparagraph{Parameters Study} Table~\ref{tab:small_parameter_learning_normalized} shows how the performance of our index varies under different parameter settings. Here, we only report the results on the \POI dataset, while the results for other datasets and the corresponding analysis are provided in the appendix (see Section~\ref{sec:ps}).
The baseline is set with $\delta = 1 \times 10^{-4}$ and $l = 5$. Specifically, $t^1$ denotes the construction time, $t^1_{k\text{NN}}$ denotes the runtime of $k$NN, and $t^1_{\text{RS}}$ denotes the runtime of radius search. For \learn with different parameters, the construction time, $k$NN runtime, and radius search runtime are denoted as $t^0$, $t^0_{k\text{NN}}$, and $t^0_{\text{RS}}$, respectively.

\finding We chose $\delta = 0.01$ because it offers a good balance between index construction and query efficiency, though not necessarily the optimal point for both. As the sample size increases, construction time first decreases due to more accurate pivot prediction and better-balanced trees, but later increases because of the higher overhead of processing larger samples. Query performance shows a similar trend: efficiency improves with more samples until the tree is sufficiently balanced, after which further increases bring little benefit. Moreover, we selected 100 sub-models as they offer a reasonable trade-off between index construction efficiency and query performance, even though this may not represent the optimal setting for both. As the number of sub-models increases, construction time first decreases due to more accurate pivot prediction and thus a more balanced tree, but then increases once the index is sufficiently balanced, since additional models add computational overhead without further improvement. Query efficiency follows the same trend: it improves as the index becomes more balanced, but eventually stabilizes.

\begin{table*}[]
\setlength{\tabcolsep}{4.2pt} 
\caption{Impact of parameters on Insertion.} 
\vspace{-0.3em}
\label{tab:insertion} 
\begin{tabular}{cccccccccccccccc}
\toprule
\multirow{2}{*}{\textbf{Dataset}} & \multicolumn{7}{c}{Sampling Rate $\delta$}                                                                                                           & \textbf{} & \multicolumn{7}{c}{\# Models $l$}               \\ \cmidrule{2-8} \cmidrule{10-16} 
                                  & $5\times10^{-4}$ & $1 \times 10^{-3}$ & $5\times 10^{-3}$ & $1\times 10^{-2}$ & $5\times 10^{-2}$ & $1 \times 10^{-1}$ & $5 \times 10^{-1}$ &           & 10   & 50   & 100  & 500  & 1000 & 5000 & 10000 \\ \midrule
\POI        & 1.37 & 1.83 & 1.85 & 1.86 & 1.86 & 1.92 & 2.46 &   & 1.04 & 1.04 & 1.02 & 1.11 & 0.98 & 0.93 & 0.98 \\
\AVL        & 1.22 & 1.22 & 1.30 & 1.24 & 1.29 & 1.39 & 1.78 &   & 1.03 & 1.02 & 1.06 & 1.06 & 1.01 & 0.95 & 0.99 \\
\Porto      & 1.06 & 1.18 & 1.20 & 1.31 & 1.15 & 1.33 & 1.44 &   & 1.03 & 0.99 & 1.04 & 0.98 & 1.11 & 1.00 & 1.01 \\
\bottomrule
\end{tabular}
\vspace{-.7em}
\end{table*}

\begin{figure*}
	\centering \includegraphics[width=1\textwidth]{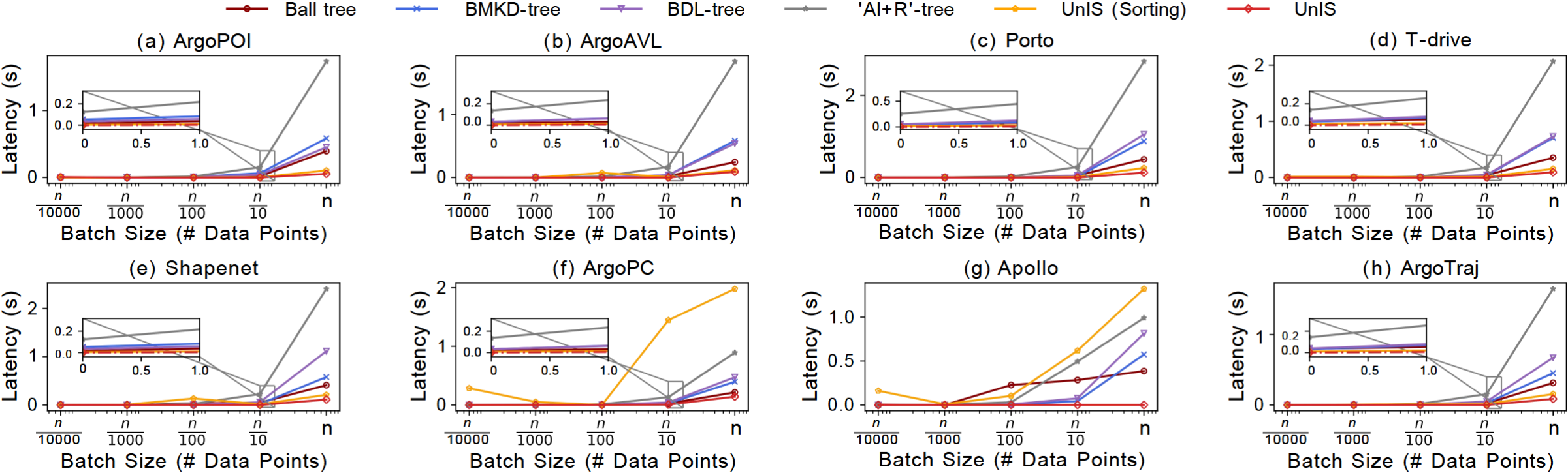}
	\vspace{-2em}
	\caption{Comparison of insertion latency of \UnIS with the SOTAs.}
	\label{fig:update}
	\vspace{-1em}
\end{figure*}

\myparagraph{Time Cost of Indexing} We study how the construction time changes with the increasing number of data points. We denote $\delta$ as the sampling rate, where $\delta \in (0,1)$. For example, with \(\delta=0.2\) and a dataset of 100 million data points, it means 20 million data points are used for index construction.
\begin{figure*}
	\centering \includegraphics[width=1\textwidth]{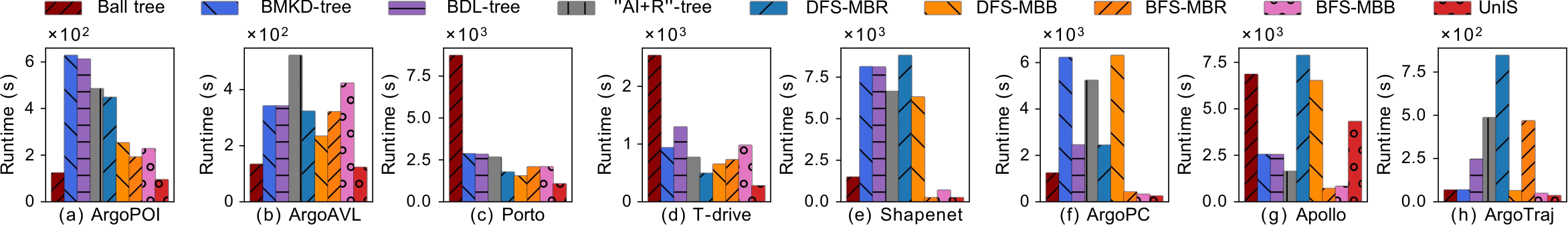}
	\vspace{-2em}
	\caption{Comparing the query runtime of \UnIS with the SOTAs.}
	\label{fig:query_vector_auto}
	\vspace{-1em}
\end{figure*}

\finding In Fig.~\ref{fig:indextime}, for most cases, the index construction time generally follows the order: \Ball $>$ \kdtree $>$ \BDLtree $>$ \aitree $>$ \UnIS. However, on a few specific datasets, \UnIS does not achieve the shortest index construction time. This is primarily because, on smaller datasets, the cost of constructing the two-stage model constitutes a larger proportion of the total indexing time.

\myparagraph{Memory Cost} We present a comparison of the memory usage of selected indexes over five datasets. In particular, ``memory usage'' represents the amount of memory occupied by Java during the index construction process.

\finding In Fig. \ref{fig:memory1}, the number of data points affects memory usage; larger datasets consume more memory. Compared with the \kdtree, our method requires slightly more memory because of the additional cost of constructing the two-stage linear regression model. However, this overhead is negligible relative to the dataset size.


\subsection{Index Insertion} 
We first conduct a parameter study to examine the impact of $\delta$ and $l$ on insertion performance, and then verify that \learn achieves lower insertion latency than SOTAs.
\label{sec:insertion}

\myparagraph{Parameter Study}
Table~\ref{tab:insertion} presents the insertion latency of our index under various parameter settings. Due to page constraints, the experiments are conducted on the \POI, \AVL, and \Porto datasets.

\finding Increasing $\delta$ initially reduces insertion latency and then causes it to rise. Small samples yield inaccurate pivot predictions and unbalanced sub-trees, while larger samples reduce tree depth, lowering reconstruction cost. Beyond a certain point, further increases provide little improvement in balance, and the overhead of handling more samples dominates, increasing latency. Similarly, as the number of sub-models grows, sub-tree rebuilding latency first decreases due to more accurate pivot predictions and better balance, then rises as additional models add overhead without further benefit.

\begin{table*}
\centering
\setlength{\tabcolsep}{6.7pt}
\ra{1.1}
\caption{Performance of the proposed feature extraction method.}
\vspace{-0.5em}
\label{tab:extraction}
\scalebox{1}{
\begin{tabular}{cccclccclccclcccc} 
\toprule
\multirow{2}{*}{\textbf{Accuracy}} & \multicolumn{3}{c}{\Unik}                                       & \multicolumn{1}{c}{} & \multicolumn{3}{c}{\aitree}                                   &  & \multicolumn{3}{c}{\TRIO}                               & \multicolumn{1}{c}{} & \multicolumn{3}{c}{\learn}                                  & \multirow{2}{*}{\textbf{$\times$Speedup}}  \\ 
\cmidrule{2-4}\cmidrule{6-8}\cmidrule{10-12}\cmidrule{14-16}
                          & $\mathbf{F}_1$ & $\mathbf{F}$ & $\mathbf{F}$-$\mathbf{F}_1$                              &                      & $\mathbf{F}_1$ & $\mathbf{F}$ & $\mathbf{F}$-$\mathbf{F}_1$                          &  & $\mathbf{F}_1$ & $\mathbf{F}$ & $\mathbf{F}$-$\mathbf{F}_1$                         &                      & $\mathbf{F}_1$ & $\mathbf{F}$ & $\mathbf{F}$-$\mathbf{F}_1$                       &                                    \\ 
\midrule
$MRR$                     & 94.87 & 95.27     & {\cellcolor[rgb]{0.753,0.753,0.753}}0.40  &                      & 87.52 & 88.51     & {\cellcolor[rgb]{0.753,0.753,0.753}}0.99 &  & 85.82 & 87.43     & {\cellcolor[rgb]{0.753,0.753,0.753}}1.61 &                      & 92.35 & 93.66     & {\cellcolor[rgb]{0.753,0.753,0.753}} 1.31& 1.08                                    \\
$\mathcal{P}_1$           & 56.2  & 56.47     & {\cellcolor[rgb]{0.753,0.753,0.753}}0.27 &                      & 57.6  & 63.38     & {\cellcolor[rgb]{0.753,0.753,0.753}} 5.78&  & 79.41 & 82.99     & {\cellcolor[rgb]{0.753,0.753,0.753}} 3.58&                      & 81.62 & 83.24     & {\cellcolor[rgb]{0.753,0.753,0.753}} 1.62&   2.81                                \\
$\mathcal{P}_2$           & 89.8  & 90.6      & {\cellcolor[rgb]{0.753,0.753,0.753}}  2.44   &                      & 80.42 & 81.56     & {\cellcolor[rgb]{0.753,0.753,0.753}}1.14 &  & 81.67 & 85.14     & {\cellcolor[rgb]{0.753,0.753,0.753}}3.47 &                      & 86.59 & 89.02     & {\cellcolor[rgb]{0.753,0.753,0.753}} 2.44&  2.37                                  \\
\bottomrule
\end{tabular}}
\vspace{-0.5em}
\end{table*}

\myparagraph{Insertion Latency} Our focus is on the latency of inserting a fixed number of data points. 
We investigate the variation in latency as the volume of inserted data points increases.
\begin{figure*}
	\centering \includegraphics[width=1\textwidth]{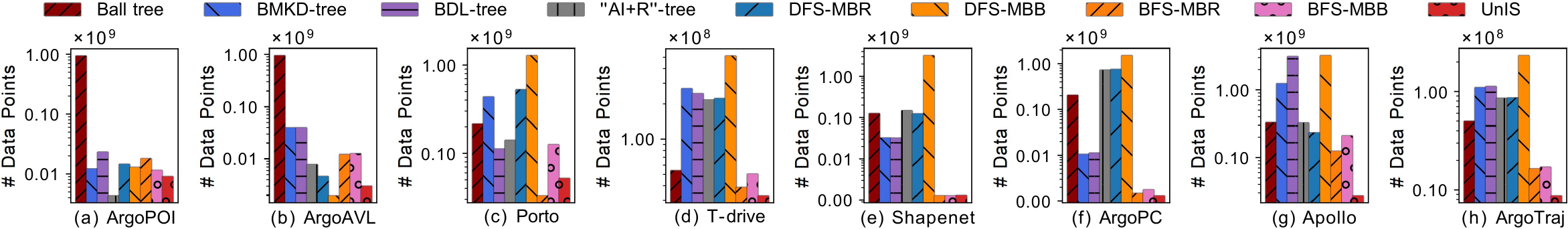}
	\vspace{-2em}
	\caption{The performance of $k$NN search of different indexes and search strategy in terms of \# data points access.}
	\label{fig:vector_traveral}
	\vspace{-1.5em}
\end{figure*}

\finding In Fig.~\ref{fig:update}, when the number of data points to be inserted is less than $\frac{n}{100}$, there is a minimal disparity in insertion efficiency between \learn and the baselines. However, as the number of data points increases, the insertion efficiency of \learn outperforms SOTAs.  


\subsection{Evaluation of Auto-selection Model}
\label{sec:auto_selection}
We first evaluate the auto-selection model and then verify that it can accelerate queries. We verify its effectiveness in accelerating $\mathrm{k}$-means in the appendix (see Section~\ref{sec:kmeans}).

\myparagraph{Impact of Key Factors} We focus on two factors, traversal methods and bounding volumes, to examine their impact on query performance, as shown in Table~\ref{tab:BDMM}. Due to space limitations, we use $k$NN as a representative query. Notably, $\frac{t_{\text{BFS}}}{t_{\text{DFS}}}$ represents the ratio of the $k$NN query time using the BFS strategy to that using the DFS strategy, while $\frac{t_{\text{MBR}}}{t_{\text{MBB}}}$ represents the ratio of the $k$NN query time using the MBR strategy to that using the MBB strategy.

\finding We found that across different datasets, traversal methods have a greater impact than minimum bounding volumes in $k$NN. This may be because, when the dataset is large and dense, adjacent nodes overlap heavily or lie very close to each other, making it difficult to prune branches or data points. In contrast, traversal methods determine the order in which data points are explored. If traversal begins near the query point, pruning becomes more effective because a tighter bound can be established earlier.

\myparagraph{Ground Truth Generation}We divide $1 \times 10^5$ query samples into three parts: 80\% for training, 10\% for validation, and 10\% for testing. Precisely, given a dataset $\mathbf{D}$ compressed by the MBR, denoted as $R = (\mathbf{lb},\mathbf{ub})$, for $k$NN search, each query involves a query point $\mathbf{x}^*$ and an integer $k$. We randomly select a data point $\mathbf{x}^*$ from $\mathbf{D}$, and $k$ is chosen randomly from the range 1 to 1000. Similarly, for radius search, each query comprises a point $\mathbf{p}^*$ and a radius $r$. We randomly select a point $\mathbf{p}^*$ from $\mathbf{D}$, and compute $r$ as $r = \sum_{i=1}^d (ub_i - lb_i)^2 \times \tau$, where $\tau$ is randomly selected from the interval [0, 1].

\myparagraph{Verification for Feature Extraction}
We evaluate the performance of the auto-selection model using mean reciprocal rank ($MRR$) \cite{Wang2020}, macro-averaged precision ($\mathcal{P}_{1}$), and micro-averaged precision ($\mathcal{P}_{2}$). The results are reported in Table~\ref{tab:extraction}.

\finding (1) The addition of features extracted by our feature extraction method results in improved predictive accuracy in most cases, as evaluated by different metrics. However, not all additional features improve prediction accuracy, as the extracted features may not always contain useful information. (2) In most cases, using our auto-selection model to predict the effective search strategy yields the highest prediction accuracy.
\begin{table}[]
\setlength{\tabcolsep}{9pt}
\ra{1.1}
\caption{Radius search performance.}
\label{tab:radius}
\vspace{-1em}
\begin{tabular}{ccccc}
\toprule
\textbf{Dataset} & \textbf{Strategy} & \textbf{Percent} & \textbf{Prediction} & \textbf{Speedup} \\ \midrule
 \POI       &        \texttt{MBR-DFS}         &   97.78\%      &  0.01\%    &  28.96\% \\
 \AVL       &       \texttt{MBR-DFS}           &    96.74\%     &    10.57\%  &  37.09\%  \\
 \Porto       &     \texttt{MBR-BFS}             &   99.54\%      &   0.02\%  &  30.92\%   \\
  \Shapenet       &   \texttt{MBR-BFS}               &    97.77\%     &  0.02\%    &  33.99\%  \\
 \Tdrive       &    \texttt{MBR-BFS}              &   97.74\%      &   0.01\%  &  30.53\%   \\
 \PC       &      \texttt{MBR-DFS}            &    98.04\%     &   18.98\%   & 36.76\%   \\
  \Apollo      &    \texttt{MBR-DFS}              &    97.87\%     &  28.96\%  &  25.59\%    \\ 
  \Trajectory       &   \texttt{MBR-DFS}               &   98.67\%      &   12.17\%  &  36.06\%   \\
  \bottomrule
\end{tabular}
\vspace{-1.7em}
\end{table}

\myparagraph{$k$NN and Radius Search Performance} The runtime of a query consists of the query execution time and the prediction time. Additional costs from index construction and periodic insertions are excluded, as construction is performed offline. Including insertion costs would introduce variability across different indexes, which could distort an accurate comparison of the intrinsic query performance of each method. For $k$NN search, the evaluation metrics are the runtime and the number of data point accesses, both reported as averages per query. For radius search, certain strategies dominate as the fastest in several datasets, causing the prediction model to consistently favor a single strategy. However, it still serves as an effective means to eliminate inefficient strategies. In Table~\ref{tab:radius}, we report the most frequently selected strategy across a set of radius search tasks (denoted as ``Strategy''), its selection ratio (denoted as ``Percent''), and the proportion of prediction time in the total query time (denoted as ``Prediction''). Moreover, we also report the corresponding time improvement relative to the average runtime of all search strategies. Notably, the runtime of a single query consists of two recurring components: query execution time and prediction time. The following experiments do not consider index construction, which can be performed offline, nor insertion performance, since insertion latency would hinder the evaluation of \learn’s query efficiency.

\finding (1) Fig.~\ref{fig:query_vector_auto} shows our auto-selection model improves $k$NN efficiency. However, it does not achieve the best performance all the time. For instance, in \PC, the model is slower than using \texttt{BFS-MBB}. This is because the minor performance gains are offset by the additional time required for prediction. (2) Fig.~\ref{fig:vector_traveral} shows that the search strategies chosen by the auto-selection model traverse the fewest points. However, this differs from Fig.~\ref{fig:query_vector_auto} because fewer traversed points do not mean shorter running time, as the total time includes both query and prediction times. (3) In Table~\ref{tab:radius}, when the most frequently selected strategy dominates with a significantly higher proportion, \learn still improves search efficiency compared to the average runtime of all strategies.




\section{Conclusions and Future Work}
\label{sec:conclusion}
In this paper, we presented \UnIS for efficient on-device search. We first designed a partition method and a lightweight ML model to improve the BMKD-tree construction efficiency. We then proposed a real-time in-place insertion operation to enable queries to access continuously generated data points. Furthermore, we proposed an auto-selection model to improve search efficiency. Experiments showed that \UnIS effectively improved the efficiency of indexing, insertion, and on-device search. In future work, we plan to extend \learn to a parallel framework to better use computational resources. We also investigate which dataset distributions benefit most from the auto-selection model in accelerating on-device search.

\section*{Acknowledgments}
This work was supported by the National Key R\&D Program of China (2023YFB4503600), National Natural Science Foundation of China (62202338, 62372337), and the Key R\&D Program of Hubei Province (2023BAB081).

\section*{AI-Generated Content Acknowledgement}
We used ChatGPT only to check grammar issues in the manuscript. All technical ideas, algorithmic designs, experimental analyses, manuscript writing, and code were entirely completed by the authors.


%





\ifCLASSOPTIONcaptionsoff
  \newpage
\fi

\bibliographystyle{abbrv}%
\bibliography{reference.bib}
\clearpage
\begin{table*}[]
\setlength{\tabcolsep}{1pt}
\caption{The accuracy of our two-stage linear model.}
\label{tab:accuracy}
\vspace{-0.5em}
\scalebox{0.805}{
\begin{tabular}{ccccccccccccccccccccccccccccc}
\toprule
\multirow{2}{*}{\textbf{
Accuracy}} & \multicolumn{2}{c}{\textbf{\POI}} &  & \multicolumn{2}{c}{\textbf{\AVL}} &  & \multicolumn{2}{c}{\textbf{\Porto}} &  & \multicolumn{2}{c}{\textbf{\Tdrive}} &  & \multicolumn{3}{c}{\textbf{\Shapenet}} &  & \multicolumn{3}{c}{\textbf{\PC}} &  & \multicolumn{3}{c}{\textbf{\Apollo}} &  & \multicolumn{4}{c}{\textbf{\Trajectory}} \\ \cmidrule{2-29} 
                                     & 1th                     & 2th                    &  & 1th                     & 2th                    &           & 1th                      & 2th                     &           & 1th                       & 2th                       &           & 1th                      & 2th              & 3th                 &           & 1th            & 2th            & 3th           &           & 1th             & 2th             & 3th             &           & 1th          & 2th          & 3th         & 4th         \\ \midrule
$r$ & 0 & 2.01E-10 &  & 3.58E-07 & 1.01E-06 &  & 1.22E-10 & 5.78E-10 &  & 1.53E-09 & 2.75E-10 &  & 4.70E-10 & 6.30E-10 & 0&  & 5.04E-08 & 3.10E-09 & 0 &  & 1.06E-07 & 3.21E-08 & 0 &  & 4.71E-08 & 3.73E-07 & 3.60E-07 & 4.64E-07 \\ \bottomrule
\end{tabular}}
\vspace{-1em}
\end{table*}

\begin{table*}[] 
\setlength{\tabcolsep}{2.7pt} 
\caption{Impact of parameters on \UnIS.} 
\vspace{-1em}
\label{tab:parameter_learning_normalized} 
\begin{tabular}{ccccccccccccccccc} 
\toprule 
\multirow{2}{*}{\textbf{Dataset}} & \multirow{2}{*}{\textbf{Parameters}} & \multicolumn{7}{c}{Sampling Rate $\delta$} & \textbf{} & \multicolumn{7}{c}{\# Models $l$} \\ 
\cmidrule{3-9} \cmidrule{11-17} 
& & $5\times10^{-4}$ & $1 \times 10^{-3}$ & $5\times 10^{-3}$ & $1\times 10^{-2}$ & $5\times 10^{-2}$ & $1 \times 10^{-1}$ & $5 \times 10^{-1}$ & & 10 & 50 & 100 & 500 & 1000 & 5000 & 10000 \\ 
\toprule 
\multirow{3}{*}{\POI} 
& $\frac{t^0}{t^1}$ & 0.70 & 0.70 & 0.61 & 0.57 & 0.55 & 1.61 & 1.75 & & 0.49 & 0.48 & 0.47 & 0.48 & 0.45 & 0.46 & 0.45 \\ 

& $t_{k\text{NN}}^0/t_{k\text{NN}}^1$ & 1.00 & 0.83 & 0.77 & 0.77 &0.77 & 0.77 &0.77 & & 0.87 & 0.87 & 0.87 & 0.87 & 0.87 &0.87 & 0.87 \vspace{0.3em}\\ 
& $t_{\text{RS}}^0/t_{\text{RS}}^0$ & 0.75 & 0.75 & 0.73 & 0.73 & 0.73 & 0.73& 0.73 & & 2.72 & 2.49 & 0.99 & 0.99 & 0.99  & 0.99  & 0.99 \\ \midrule

\multirow{3}{*}{\AVL} 
& $\frac{t^0}{t^1}$ & 0.79 & 0.68 & 0.79 & 0.81 & 0.68 & 1.65 & 1.77 & & 0.53 & 0.55 & 0.63 & 0.56 & 0.40 & 0.35 & 0.38 \\

& $t_{k\text{NN}}^0/t_{k\text{NN}}^1$ & 1.21 & 1.13 & 1.17 & 0.91 & 0.63 & 0.63 & 0.65 & & 0.92 & 0.92 &0.92 & 0.92 & 1.00 & 1.01 & 0.94 \vspace{0.3em} \\ 
& $t_{\text{RS}}^0/t_{\text{RS}}^0$ & 0.90 & 0.91 & 0.91 & 0.91 & 0.91& 0.91 & 0.91 & & 1.00 &1.00 & 0.99 & 0.99 & 0.99 &0.99& 0.99 \\ \midrule

\multirow{3}{*}{\Porto} 
& $\frac{t^0}{t^1}$ & 0.81 & 0.67 & 0.63 & 0.55 & 0.57 & 0.54 & 0.65 & & 0.45 & 0.45 & 0.42 & 0.40 & 0.40 & 0.37 & 0.37 \\ 
& $t_{k\text{NN}}^0/t_{k\text{NN}}^1$ & 0.89 & 0.84 & 0.83 & 0.89 & 0.87 & 0.89 & 0.87 & & 1.00 & 1.00 & 1.00 & 1.00 & 1.00 & 1.00 & 1.00 \vspace{0.3em} \\ 
& $t_{\text{RS}}^0/t_{\text{RS}}^0$ & 0.96 & 1.17 & 0.99& 0.97 & 0.97 & 0.97& 0.97 & & 0.89 &  0.89 &  0.89&  0.89&  0.89 & 0.70 & 0.74 \\ \midrule

\multirow{3}{*}{\Shapenet} 
& $\frac{t^0}{t^1}$ & 0.97 & 0.74 & 0.67 & 0.61 & 0.58 & 0.64 & 0.92 & & 0.58 & 0.53 & 0.53 & 0.54 & 0.48 & 0.63 & 0.66 \\

& $t_{k\text{NN}}^0/t_{k\text{NN}}^1$ & 1.00 & 1.00 & 1.00 & 1.00 & 1.00 & 1.00 & 1.00& & 0.90 & 0.90 & 0.90 &0.90& 0.88 & 0.88& 0.88 \vspace{0.3em} \\ 
& $t_{\text{RS}}^0/t_{\text{RS}}^1$ & 0.98 & 0.98 & 0.98 & 0.98 & 0.96 & 0.96 & 0.96 & & 1.00 & 1.00 & 1.00 & 1.00 & 1.00 &0.99 & 0.99 \\ \midrule

\multirow{3}{*}{\Tdrive} 
& $\frac{t^0}{t^1}$ & 0.86 & 0.67 & 0.59 & 0.55 & 0.51 & 0.44 & 0.47 & & 0.33 & 0.31 & 0.27 & 0.27 & 0.21 & 0.17 & 0.19 \\ 

& $t_{k\text{NN}}^0/t_{k\text{NN}}^1$ & 1.00 & 1.00 & 1.00  & 1.00  & 1.00  & 0.00 & 1.00  & & 0.98  & 0.97 & 0.98 & 0.97 & 0.97 & 0.97 & 0.97 \vspace{0.3em} \\ 
& $t_{\text{RS}}^0/t_{\text{RS}}^1$ & 0.92 & 0.92 & 0.83 & 0.83&0.83& 0.83& 0.83& & 0.93 & 0.93 & 0.93 & 0.93 & 0.93  & 0.93 & 0.93  \\  \midrule

\multirow{3}{*}{\PC} 
& $\frac{t^0}{t^1}$ & 0.90 & 0.93 & 0.84 & 0.73 & 0.71 & 0.73 & 1.07 & & 0.51 & 0.50 & 0.47 & 0.34 & 0.31 & 0.39 & 0.42 \\ 

& $t_{k\text{NN}}^0/t_{k\text{NN}}^1$ & 0.78 & 0.76 &0.78 & 0.78 & 0.78 &0.78 &0.78 & & 0.99 &0.99  & 0.99  &0.99  & 0.99 &0.99  & 0.99  \vspace{0.3em}\\ 
& $t_{\text{RS}}^0/t_{\text{RS}}^1$  & 0.92 & 0.92 & 0.92 & 0.89& 0.89 & 0.89 & 0.90 & &0.93 & 0.93 & 0.92& 0.92 & 0.92 &0.92 &0.92\\  \midrule

\multirow{3}{*}{\Apollo} 
& $\frac{t^0}{t^1}$ & 0.78 & 0.77 & 0.69 & 0.68 & 0.57 & 0.67 & 0.78 & & 0.53 & 0.47 & 0.41 & 0.43 & 0.39 & 0.28 & 0.34 \\ 

&  $t_{k\text{NN}}^0/t_{k\text{NN}}^1$ & 0.92 & 0.90 & 0.90 & 0.90& 0.91 & 0.89 & 0.89 & & 0.99 & 0.99 & 0.99 &0.97 & 0.97 & 0.97 & 0.97\vspace{0.3em}  \\ 
& $t_{\text{RS}}^0/t_{\text{RS}}^1$ & 0.88 &  0.86& 0.85 & 0.87 & 0.85&  0.84 & 0.87 & & 0.92 & 0.92 & 0.92 & 0.91&0.91 & 0.91 &0.90\\  \midrule

\multirow{3}{*}{\Trajectory} 
& $\frac{t^0}{t^1}$ & 0.89 & 0.75 & 0.72 & 0.61 & 0.63 & 0.51 & 0.64 & & 0.44 & 0.41 & 0.39 & 0.38 & 0.35 & 0.34 & 0.38 \\ 

&  $t_{k\text{NN}}^0/t_{k\text{NN}}^1$  & 0.87 & 0.87 & 0.87 & 0.87 & 0.87 & 0.87 & 0.87 & & 0.90 & 0.82 & 0.82 & 0.86 & 0.86 & 0.82& 0.85 \vspace{0.3em} \\ 
& $t_{\text{RS}}^0/t_{\text{RS}}^1$ & 1.00  & 1.00 & 1.00 &1.00  & 1.00 & 1.00  & 1.00 & & 1.00  &1.00 & 1.00  & 1.00  &1.00 &1.00  & 1.00  \\ 
\bottomrule
\end{tabular} 
\end{table*}

\section{Appendix}
\label{sec:appendix}
\subsection{Radius Search}
\label{sec:rs}
Algorithm~\ref{alg:range_search_total} describes the radius search procedure using DFS (line~\ref{line:rs_call_dfs}) or BFS (line~\ref{line:rs_call_bfs}). In DFS, the algorithm recursively explores nodes, pruning subtrees using Lemma \ref{lemma:Ball_filter} (line~\ref{line:rs_prune_dfs}). 
At the leaves (line~\ref{line:rs_leaf_dfs}), candidate points are checked, and those within the radius are added to the result set (line~\ref{line:rs_add_dfs}).
Traversal continues recursively with child nodes (lines~\ref{line:rs_recurse_left}–\ref{line:rs_recurse_right}). In BFS, nodes are processed level by level: the root node is enqueued (line~\ref{line:bfs_enqueue_root}), nodes are expanded iteratively (lines~\ref{line:bfs_loop}–\ref{line:bfs_dequeue}), subtrees are pruned when possible using Lemma \ref{lemma:intersect} (line~\ref{line:bfs_prune}), leaf points are tested and added to the result set (lines~\ref{line:bfs_leaf}–\ref{line:bfs_add}), and non-leaf children are enqueued for further exploration (lines~\ref{line:bfs_enqueue_left}–\ref{line:bfs_enqueue_right}).
\begin{algorithm}\small
\caption{\texttt{Radius\_Search}($Q$, $T$)}
\KwIn{$Q$: query ball,  $T$: traversal strategy.}
\KwOut{$P$: result set.}
\label{alg:range_search_total}
$P \gets \emptyset$, $N \gets$ the root node\;
\eIf{$T$ is depth-first search}{
    \textsf{DFS}($Q$, $N$, $P$)\; \label{line:rs_call_dfs}
}{
    \textsf{BFS}($Q$, $N$, $P$)\; \label{line:rs_call_bfs}
}
\KwRet $P$\;
\vspace{-0.5em}
\algrule
\vspace{-0.5em}

\SetKwFunction{FDFS}{\textsf{DFS}\!}
\SetKwProg{Fn}{Function}{}{}
\Fn{\FDFS{$Q$, $N$, $P$}}{
    \If{$N$ is pruned by is pruned by Lemma 2}{\Return\;\label{line:rs_prune_dfs}}
    \If{$N$ is leaf}{ \label{line:rs_leaf_dfs}
        \ForEach{$p \in N$}{ 
            \If{$p \in Q$}{$P \gets P \cup \{p\}$\;\label{line:rs_add_dfs}}
        }
        \Return\;
    }
    \If{$N.left \neq \emptyset$}{
        \textsf{DFS}($Q$, $N.left$, $P$)\;\label{line:rs_recurse_left}
    }
    \If{$N.right \neq \emptyset$}{
        \textsf{DFS}($Q$, $N.right$, $P$)\;\label{line:rs_recurse_right}
    }
}
\vspace{-0.5em}
\algrule
\vspace{-0.5em}

\SetKwFunction{FBFS}{\textsf{BFS}\!}
\SetKwProg{Fn}{Function}{}{}
\Fn{\FBFS{$Q$, $N$, $P$}}{
    $Q \gets$ empty queue\; \label{line:bfs_init}
    $Q.add(N)$\; \label{line:bfs_enqueue_root}
    \While{$Q$ is not empty}{ \label{line:bfs_loop}
        $U \gets Q.poll()$\; \label{line:bfs_dequeue}
        \If{$U$ is pruned by Lemma 1}{\textbf{continue}\;\label{line:bfs_prune}}
        \If{$U$ is leaf}{ \label{line:bfs_leaf}
            \ForEach{$p \in U$}{
                \If{$p \in Q$}{$P \gets P \cup \{p\}$\;\label{line:bfs_add}}
            }
        }
        \Else{
            \If{$U.left \neq \emptyset$}{$Q.add(U.left)$\;\label{line:bfs_enqueue_left}}
            \If{$U.right \neq \emptyset$}{$Q.add(U.right)$\;\label{line:bfs_enqueue_right}}
        }
    }
}
\end{algorithm}

\begin{figure}
	\centering
      \includegraphics[width=0.47\textwidth]{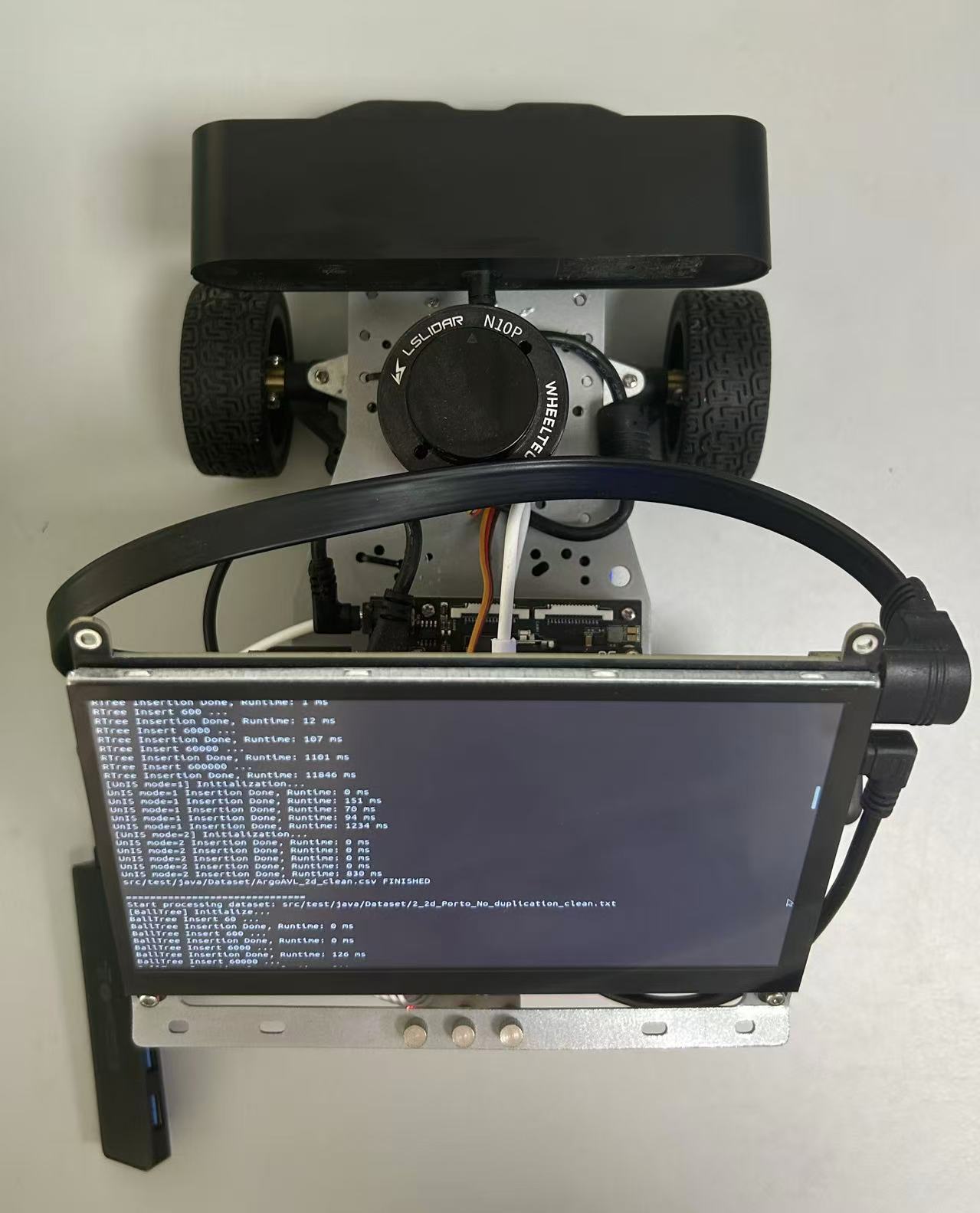}
	\vspace{-1em}
 \caption{Deployment of \learn on Wheeltec R550.}
	\label{fig:oppo_running}
    \vspace{-1em}
\end{figure}

\subsection{Wheeltec R550}
\label{sec:car}
The R550 (shown in Fig.~\ref{fig:oppo_running}) is a wheeled mobile robot with multiple chassis options, equipped with cost-effective sensors such as radar and depth cameras, making it suitable for learning applications in mapping and navigation, deep learning, 3D vision, and multi-robot coordination. Notably, it is powered by a Jetson Nano \cite{Nano} as the main processor, featuring a 64-bit ARM Cortex-A57@1.43GHz and 4GB of memory.


\subsection{Evaluating Two-Stage Regression Model}
\label{sec:two}
There is a trade-off between quantile prediction accuracy and runtime. However, a slight reduction in accuracy can lead to substantial improvements in efficiency. Specifically, we define the prediction error as $r = |\text{actual quantile} - \text{predicted quantile}|$. A larger $r$ indicates lower accuracy, while a smaller $r$ corresponds to higher accuracy. For each dataset, we extract each dimension as an array and predict the median of each array to evaluate the accuracy of our method. Details are discussed below. 

\finding As shown in Table~\ref{tab:accuracy}, our two-stage linear model achieves a prediction error $r$ of less than 1\%. In some cases, the model can directly predict the exact position of the median; for example, in the first dimension of the \POI dataset treated as an array, the prediction error is 0.

\subsection{Parameter Study}
\label{sec:ps}
The goal of our experiments was not to identify a universally applicable parameter setting, but rather to examine how the performance of our index changes with parameter variation. As shown in Table~\ref{tab:parameter_learning_normalized}, the baseline is defined with $\delta = 1 \times 10^{-4}$ and $l = 5$. Specifically, $t^1$ denotes the construction time, $t^1_{k\text{NN}}$ the runtime of $k$NN search, and $t^1_{\text{RS}}$ the runtime of radius search. For \learn built with different parameters, the corresponding times are denoted as $t^0$, $t^0_{k\text{NN}}$, and $t^0_{\text{RS}}$, respectively. A detailed discussion is provided below.

\finding (1) First, as the number of sampled points increases, the index construction time initially decreases and then increases. When the sampling rate is small (e.g., $10^{-4}$), pivot prediction is less accurate, yielding a less balanced index. With larger samples, prediction improves, producing a more balanced index with reduced depth and lower computation cost. However, when the sampling rate continues to increase, the index balance no longer improves significantly, while the computational overhead of handling larger samples increases, which in turn raises the overall construction time. (2) As the sampling rate increases, query efficiency generally improves. However, a larger sample size does not always guarantee higher search efficiency, since once the tree becomes sufficiently balanced.
(3) As the number of sub-models increases, index construction time first decreases and then increases. With fewer sub-models (e.g., 10), pivots are predicted less accurately, producing a less balanced index. As the number grows, pivot prediction improves, the index becomes more balanced with fewer levels, and construction cost decreases. However, once the index is sufficiently balanced, further increases add computational overhead without improving balance, which raises construction time. (4) As the number of sub-models increases, query efficiency generally improves. However, once the tree is balanced, adding more sub-models does not further reduce query time, which stabilizes within a certain range.

\begin{figure*}
	\centering \includegraphics[width=1\textwidth]{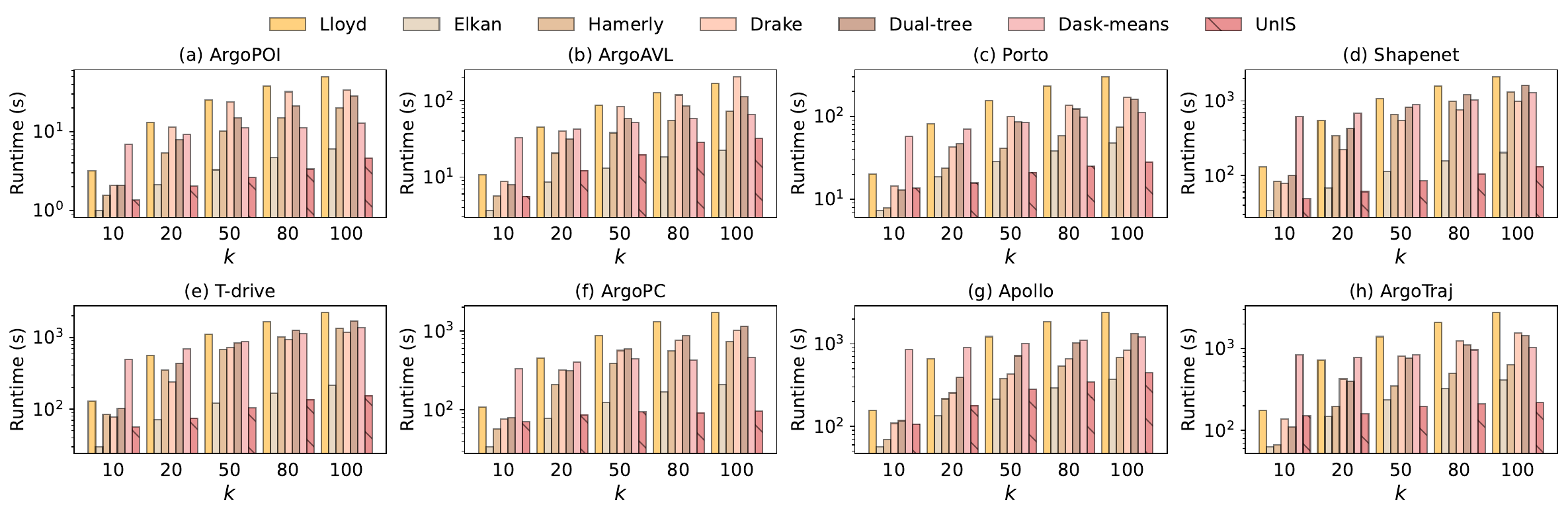}
	\vspace{-2.3em}
	\caption{Runtime performance of \UnIS when applied to the $\mathrm{k}$-means algorithm.}
	\label{fig:runtime}
	\vspace{-2em}
\end{figure*}

\subsection{Verification for $\mathrm{k}$-means}
\label{sec:kmeans}
We demonstrate the effectiveness of \UnIS when applied to $\mathrm{k}$-means by comparing its runtime with other algorithms on edge devices. We follow the approach in \cite{Yushuai2024}, which proposes using two tree-based indexes to store data points and nodes, and leverages $k$NN to avoid unnecessary distance computations. Specifically, we evaluate performance using different values of $k \in \{10, 20, 50, 80, 100\}$

\finding As shown in Fig.~\ref{fig:runtime}, when $\mathrm{k}$ is small (e.g., $\mathrm{k}=10$), \UnIS performs better than most SOTAs in pruning in most cases, but it's not always the best. For example, \Elkan outperforms it because \UnIS requires additional time to construct our indexes, while $k$NN on these indexes is inefficient when \( \mathrm{k} \) is small. Whereas when $\mathrm{k}$ is large, \UnIS demonstrates superior runtime performance due to the effective pruning power of the centroid index. 
\end{document}